\documentclass{article}
\usepackage[a4paper,margin=2.5cm]{geometry}
\usepackage[utf8]{inputenc}
\usepackage{amsmath}
\usepackage{amsfonts}
\usepackage{amssymb}
\usepackage{amsthm}

\usepackage{orcidlink}

\usepackage{mathtools}
\usepackage{thmtools}

\usepackage{hyperref}
\usepackage{cleveref}

\usepackage{subcaption}

\usepackage{setspace}
\usepackage{graphicx} 
\usepackage{pgfplots}
\usepgfplotslibrary{groupplots}
\pgfplotsset{compat=1.18}
\usepackage{tikz}
\usetikzlibrary{shapes.geometric, arrows.meta}
\usepackage{url}
\usepackage{booktabs}
\usepackage{here}
\usepackage{tabularx}
\usepackage{array}

\newcommand{\datpath}{dat/}

\definecolor{viridisViolet}{RGB}{68,1,84}
\definecolor{viridisBlue}{RGB}{59,82,139}
\definecolor{viridisTeal}{RGB}{33,145,140}
\definecolor{viridisGreen}{RGB}{94,201,98}
\definecolor{viridisYellow}{RGB}{253,231,37}
\definecolor{matchingRed}{RGB}{208,31,60}
\definecolor{matchingOrange}{RGB}{255,192,119}
\definecolor{refGray}{RGB}{120,120,120}

\pgfplotsset{gmmMarg/.style={color=matchingRed, solid, mark=*}}
\pgfplotsset{gmmKin/.style={color=matchingOrange, dashed, mark=triangle*}}
\pgfplotsset{gmmSent/.style={color=viridisViolet, solid, mark=square*}}
\pgfplotsset{gmmMean/.style={color=viridisViolet, dotted, mark=square*}}
\pgfplotsset{gmmLOCF/.style={color=viridisViolet, densely dashed, mark=square*}}
\pgfplotsset{gbmSent/.style={color=viridisTeal, solid, mark=star}}
\pgfplotsset{gbmMean/.style={color=viridisBlue, solid, mark=o}}
\pgfplotsset{gbmLOCF/.style={color=viridisGreen, solid, mark=diamond*}}
\pgfplotsset{gbmSentClean/.style={color=viridisTeal, dashed, mark=star}}
\pgfplotsset{gbmSentAug/.style={color=viridisTeal, solid, mark=star}}
\pgfplotsset{gbmMeanClean/.style={color=viridisBlue, dashed, mark=o}}
\pgfplotsset{gbmMeanAug/.style={color=viridisBlue, solid, mark=o}}
\pgfplotsset{gbmLOCFClean/.style={color=viridisGreen, dashed, mark=diamond*}}
\pgfplotsset{gbmLOCFAug/.style={color=viridisGreen, solid, mark=diamond*}}
\pgfplotsset{filterGMM/.style={color=matchingRed, solid, mark=*}}
\pgfplotsset{perFrameGBM/.style={color=viridisBlue, densely dashed, mark=diamond*}}
\pgfplotsset{filterGBM/.style={color=viridisBlue, solid, mark=o}}
\pgfplotsset{filterGbmIid/.style={color=viridisBlue, solid, mark=o}}
\pgfplotsset{filterGbmBursty/.style={color=viridisBlue, densely dashed, mark=o}}
\pgfplotsset{filterGmmIid/.style={color=matchingRed, solid, mark=*}}
\pgfplotsset{filterGmmBursty/.style={color=matchingRed, densely dashed, mark=*}}
\pgfplotsset{perFrameGbmIid/.style={color=viridisBlue, densely dashed, mark=diamond*}}
\pgfplotsset{perFrameGbmBursty/.style={color=viridisBlue, dotted, mark=diamond*}}
\pgfplotsset{stackedGBM/.style={color=viridisTeal, solid, mark=square*}}
\pgfplotsset{rule/.style={color=matchingOrange, dashed, mark=triangle*}}
\pgfplotsset{stackShort/.style={color=viridisTeal, solid, mark=square*}}
\pgfplotsset{stackLong/.style={color=viridisTeal, densely dashed, mark=square*}}
\pgfplotsset{stackIid/.style={color=viridisTeal, solid, mark=square*}}
\pgfplotsset{stackBursty/.style={color=viridisTeal, densely dashed, mark=square*}}
\pgfplotsset{floor/.style={color=refGray, dotted}}
\pgfplotsset{kinematics/.style={color=viridisViolet, solid, mark=*}}
\pgfplotsset{compact/.style={color=viridisTeal, solid, mark=triangle*}}
\pgfplotsset{fullCont/.style={color=viridisGreen, solid, mark=diamond*}}
\pgfplotsset{fullPresent/.style={color=viridisYellow, solid, mark=pentagon*}}
\pgfplotsset{gmm/.style={color=viridisTeal, solid, mark=triangle*}}
\pgfplotsset{gbm/.style={color=viridisBlue, dashed, mark=o}}
\pgfplotsset{cond/.style={color=matchingRed, dotted, mark=square*}}

\pgfplotsset{
  compat=1.18,
  paperaxis/.style={
    width=\paperfigwidth, height=\paperfigheight,
    tick align=outside, tick pos=left,
    axis line style={draw=black!55, line width=0.5pt},
    every tick/.style={black!55, line width=0.4pt},
    major tick length=2.4pt, minor tick length=1.2pt,
    grid=major, major grid style={draw=black!12, line width=0.4pt},
    label style={font=\small}, tick label style={font=\footnotesize},
    title style={font=\small\bfseries, yshift=1pt},
    legend style={font=\footnotesize, draw=black!30, fill=white,
                  fill opacity=0.92, text opacity=1, inner sep=2.4pt,
                  row sep=0.4pt, /tikz/nodes={inner sep=1.6pt}},
    legend cell align=left,
    every axis plot/.append style={line width=1.0pt, mark size=1.7pt,
                                   mark options={solid}},
  },
}
\providecommand{\paperfigwidth}{0.46\linewidth}
\providecommand{\paperfigheight}{4.2cm}

\newtheorem{theorem}{Theorem}[section]

\theoremstyle{remark}
\newtheorem{remark}[theorem]{Remark}
\Crefname{remark}{Remark}{Remarks}

\renewcommand{\P}[1]{\mathbb{P}\left[#1\right]}
\newcommand{\I}{{I}}
\renewcommand{\O}{{\mathcal{O}}}

\newcommand{\R}{\mathbb{R}}
\usepackage[dvipsnames]{xcolor}

\title{Robust lane-change intention anticipation under uncertainty based on a recursive Bayesian filtering approach\thanks{This publication is based upon work from COST Action InterCoML, supported by COST (European Cooperation in Science and Technology).}}

\author{Dilara Kılınç\,\orcidlink{0009-0001-2035-1399}\thanks{Budapest University of Technology and Economics, Budapest, Hungary, ({\tt kilinc.dilara@edu.bme.hu}).}\textsuperscript{\, ,\,}\thanks{Corresponding author, email: {\tt kilinc.dilara@edu.bme.hu}} \and Hendrik Kleikamp\,\orcidlink{0000-0003-1264-5941}\thanks{IDea\_Lab, University of Graz, Austria, ({\tt hendrik.kleikamp@uni-graz.at}).} \and Bruno Viti\,\orcidlink{0009-0002-5593-4743}\thanks{Department of Mathematics and Scientific Computing, University of Graz, Austria, ({\tt bruno.viti@uni-graz.at}).}}
\date{August 21, 2026}

\begin{document}

\maketitle

\begin{abstract}
\noindent In this paper we discuss and analyze a method for lane-change intention anticipation of drivers on highways based on kinematic features and surrounding observations.
The approach makes use of a recursive Bayesian filtering strategy, which can be interpreted as a hidden Markov model.
We present two models for the likelihood term and show how they can handle missing observations.
An additional focus of the work lies on calibrating the obtained probabilities in order to obtain reliable predictions.
The introduced approach is evaluated in practice on the highD dataset and compared to several other baseline methods.
Particular emphasize is put on evaluating the robustness and temporal consistency as well as calibration of the developed algorithm.
Extensive numerical experiments allow for a careful performance assessment and detailed discussion of advantages and limitations of the different methodologies.
\end{abstract}


\section{Introduction}
\noindent
Lane changes are among the complex and safety-critical maneuvers on highways as they require both longitudinal and lateral motion involving continuous interactions with vehicles in adjacent lanes~\cite{Cristofaro2026}. In the context of autonomous driving systems, accurately forecasting the intentions of surrounding vehicles is essential for proactive risk mitigation and for enabling safe responses to dynamic traffic conditions. While recent advances in Intelligent Transportation Systems~(ITS), including Vehicle-to-Everything~(V2X), have significantly reduced information asymmetry and perceptual uncertainties by enabling cooperative awareness, notable challenges remain in heterogeneous traffic environments~\cite{Li2025,Song2022}. In particular, the presence of human-driven vehicles, which are expected to constitute the majority on the roads for the foreseeable future, presents a major challenge. In contrast to connected autonomous vehicles, human drivers often make sudden or rapid changes, including overtaking and abrupt shifts in driving intention. These unpredictable behaviors introduce significant uncertainty into the driving environment, potentially leading to hazardous situations if not anticipated in a timely manner~\cite{Alqahtani2026}. 
\par A lane change anticipation task seeks to infer a driver's intention for a lane-change maneuver before its physical execution begins. Moreover, a driver's intention develops gradually over time and becomes apparent as the driving situation unfolds. Therefore, the anticipation task is inherently temporal and uncertain, as observations obtained at successive time steps provide continuously evolving information about future intention~\cite{Li2022}. Given this temporal dependency, predictions may become more sensitive to missing or temporarily unavailable information, resulting in unstable intention estimates. Treating missing values as observed values may introduce artificial cues and lead to overconfident estimates~\cite{Li2025}. In contrast, discarding incomplete observations may eliminate early indications of the impending information that is critical for anticipation. Taken together, these considerations underscore the need to develop anticipation models that can appropriately handle missing information.
\par  It is worth noting that a model may correctly distinguish lane-changing events while producing estimates that are overconfident or underconfident. On the one hand, predictive accuracy has been the primary focus of many studies for lane change anticipation tasks, with considerable effort devoted to improving the prediction horizon. On the other hand, beyond predicting the most likely maneuver, it is also essential to consider whether the confidence associated with that prediction is reliable. This raises the question of whether the model's outputs are properly calibrated. A model is well calibrated if its predicted probabilities are consistent with the observed frequencies of the corresponding outcomes~\cite{SilvaFilho2023}. In an intention anticipation task, miscalibration may propagate through downstream planning modules, influencing how conservatively the system responds. Therefore, developing robust, reliable and trustworthy intention anticipation models is essential for enabling safe and informed decision-making in safety-critical autonomous driving applications.

\subsection{Literature review}
\noindent
This section reviews the relevant literature on anticipating lane-change intentions. It begins with an overview of traditional machine learning approaches, followed by deep learning-based approaches for maneuver prediction with particular attention to temporal consistency. We then consider studies addressing the reliability of these predictions, emphasizing probability calibration.

\subsubsection{Traditional machine learning approaches}
\noindent 
Conventional approaches to lane-change intention anticipation have explored a range of machine learning and probabilistic methods to capture driving behavior on highways. Support Vector Machines~(SVMs) are commonly used to classify likely maneuvers of human-driven vehicles~\cite{Manzour2026}. Kumar \textit{et al.}~\cite{Kumar2013} improved this approach by combining multiclass SVM outputs with Bayesian filtering to enhance temporal consistency for real-time intention anticipation on highways. An alternative approach proposed by Morris and Doshi~\cite{Morris2011} used the Relevance Vector Machine~(RVM), which is a Bayesian inference model based on~SVM. With the availability of larger naturalistic driving datasets, research has increasingly moved to data-driven approaches that can model the temporal evolution of driving behavior~\cite{https://doi.org/10.48550/arxiv.2607.09740}. Sharma \textit{et al.}~\cite{Sharma2022} proposed a hierarchical architecture that first performs binary classification of driver intentions using an SVM and then classifies the intentions with the direction of lane change using a continuous hidden Markov model combined with a Gaussian mixture model~(GMM). The model achieved high classification accuracy on real-world~NGSIM data. Gradient boosting machines~(GBM) have also been applied to capture nonlinear relationships between driving-related features and maneuver intentions~\cite{Li2022}.

\subsubsection{Deep learning approaches}
\noindent
As deep learning progresses, more advanced and comprehensive methods have been adopted to enhance lane change prediction models. In this context, convolutional neural networks~(CNNs), recurrent neural networks~(RNNs) such as long short-term memory networks~(LSTMs), gated recurrent units~(GRUs), and the increasingly sophisticated Transformer models have demonstrated exceptional performance in predicting driving intentions~\cite{Zhou2025,Cristofaro2026}. These models incorporate high-level spatio-temporal features from the historical context of previous observations, providing reliable long-term forecasts. More recent approaches have further extended these models by incorporating graph-based representations for temporal dependencies. Ren \textit{et al.}~\cite{Ren2025} explored~XGBoost and~LSTM-based models that incorporate real-time traffic context, while Geng \textit{et al.}~\cite{Geng2023} combine~XGBoost with a bidirectional gated recurrent unit~(BiGRU) to further refine the prediction of lane-changing intentions, achieving an overall accuracy of over~98\%. In a different direction, Gao \textit{et al.}~\cite{Gao2023} developed a dual-transformer architecture for intention prediction. In this sense, these approaches are conceptually aligned with the recursive filtering adopted in our study.

\subsubsection{Reliability in intention anticipation}
Uncertainty in machine learning models can arise from several factors, such as sensor quality, model training procedures, and incomplete observations, which influence both the correctness of predictions and their confidence level; predictive accuracy alone does not indicate a model's reliability~\cite{Vellenga2024}. Accordingly, in addition to performance of the predictive accuracy, researchers have increasingly focused on improving the reliability of model estimates through probability calibration~\cite{guo2017calibration}. Several approaches, such as deep ensembles, where multiple networks are trained independently, have been investigated to improve the calibration and robustness of predictive models~\cite{https://doi.org/10.48550/arxiv.1612.01474}. Along the same line, Gal \textit{et al.}~\cite{pmlr-v48-gal16} introduced dropout during inference, which can be interpreted as a variational approximation for estimating uncertainty in neural networks. Post-hoc methods, instead, such as temperature scaling, adjust model confidence using a validation set without requiring any additional model training, while training-based approaches have been investigated to reduce overconfident errors by explicitly penalizing overconfident predictions through modified training losses~\cite{guo2017calibration,wang2023calibrating}. In the autonomous driving field, Ozkan \textit{et al.}~\cite{zkan2026} proposed a physics-guided multimodal prediction model and assessed the calibration of its probability estimates using the expected calibration error~(ECE). Their calibration analysis showed that temperature scaling reduced the expected calibration error from~0.043 to~0.038, while maintaining prediction accuracy. In the lane-change intention setting, early research has shown that conventional softmax-based neural networks can produce highly confident predictions even when those predictions are incorrect or when the inputs are out of the data they were trained on~\cite{Lin2026}. Nevertheless, the calibration of lane-change intention prediction under incomplete observations has not been adequately investigated. Together, these findings highlight the need to assess the reliability of the associated probability through calibration as model predictions support safety-critical decisions.

\subsection{Overview of the paper}
\noindent 
A driver's intention develops gradually over time and becomes apparent as the driving situation unfolds. This temporal nature motivates us to formulate lane-change intention anticipation as a recursive inference problem. To capture this evolving process, we propose a recursive Bayesian filter that takes new observations into account and updates the posterior probability of each driving intention at every time step. By incorporating observations as they come in, the filter maintains an evolving belief state and provides a probabilistic estimate of the driver's intention. 
To provide the observation likelihoods required by the recursive Bayesian filter, we consider two emission models. From one perspective, we use a Gaussian mixture model~(GMM) to model the distribution of the observation vector for each driving intention. Modeling these distributions provides a principled way to handle incomplete observations through analytic marginalization. In practice, at a given time, the model can process incomplete observations without filling in missing data or retraining on corrupted data. This approach provides a robust treatment of imperfect sensing conditions caused by sensor failures, occlusions, communication delays or temporary data loss. From another perspective, as the dimensionality of the observation space increases, GMMs become more difficult to optimize. In particular, covariance structures of the Gaussian components may become less reliable, and the expectation-maximization process may become poorly conditioned in high-dimensional observation spaces. To benefit from the predictive power of discriminative classifiers, we introduce a second emission model based on a gradient boosting machine~(GBM) using a scaled-likelihood trick to estimate the probability of each of the three intention classes. The scaled-likelihood trick transforms the posterior into a likelihood term by incorporating the prior into the recursive Bayesian filter. This approach is well suited to complete feature configuration while missing features are handled through imputation. Thus, the~GMM and~GBM present different practical properties: the GMM provides a principled treatment of missing features, whereas the~GBM offers an alternative for high-dimensional observations. We next consider integrating the emission model with a recursive Bayesian filter. The proposed recursive Bayesian filter reduces the influence of short-term fluctuations that could cause unnecessary changes in the estimated intention, resulting in fewer abrupt prediction changes and a lower flip-rate. The filter consequently supports temporal consistency with a high predictive accuracy; however, the associated probabilities must also reflect reliable anticipation. We therefore apply calibration methods to improve the correspondence between the predicted probabilities and the observed frequencies of the corresponding intention classes. Finally, we present a comprehensive evaluation of the proposed framework on the~highD dataset~\cite{Krajewski2018} across multiple prediction horizons. This evaluation considers not only predictive accuracy, but also robustness to missing observations, temporal consistency, probability calibration, and computational efficiency. Taken together, we discuss detailed results including comparisons against relevant baseline methods. Overall, we provide an interpretable and transparent probabilistic representation of lane-change intention anticipation while maintaining millisecond real-time inference. 
\newline\newline
\noindent The paper is organized as follows:
We start by describing the problem statement of driver intention prediction in~\Cref{sec:problem-statement}.
Moreover, we introduce the data available in the~highD dataset and discuss how we introduce random observation dropout in order to generate more realistic data.
In~\Cref{sec:recursive-bayesian-filtering} the recursive Bayesian filtering approach is presented in detail together with different strategies for incorporating high-dimensional observation vectors, missing observations and calibration of the resulting probabilities for the intention classes.
By means of extensive experiments on the~highD dataset in~\Cref{sec:experiments}, we evaluate our approach and compare it to different baselines.
We investigate in particular the performance of our method under sensor dropout, its temporal consistency in terms of flip-rate and sustained commitment to predictions as well as training and inference runtime.
The paper ends with a comprehensive discussion of advantages and limitations of the different approaches in~\Cref{sec:discussion-outlook} together with an outlook to future lines of research.

\section{The problem of anticipating driver intentions on highways}\label{sec:problem-statement}
In a mixed traffic environment, autonomous vehicles and human-driven vehicles share the same road. Human drivers possess an inherent ability to infer the intentions of other road users through direct observation and contextual understanding. This intuitive anticipation is crucial for safe and reliable driving. However, the autonomous vehicles do not have this intuitive human understanding. Therefore, they need reliable prediction models to estimate the future intentions of surrounding vehicles. This problem becomes especially significant in highway driving scenarios. On highways, lane keeping and lane chancing are fundamental driving intentions that demand precise and timely prediction. 
\par
A lane change maneuver involves an alteration of the spatial relationship between subject vehicle and its neighbors. For instance, the distance to the front vehicle, rear vehicle, and vehicles in adjacent lanes may change during this maneuver. Therefore, early and accurate anticipation of lane change intention is necessary for safe motion planning, risk assessment, and situational awareness in autonomous driving. 
\par
A further challenge in real-world autonomous driving is that sensor data may not always be complete or reliable. Observations can be noisy, partially missing or temporarily unavailable due to sensor limitations and environmental conditions. For this reason, a model for driver intention anticipation should also be able to handle missing information.
\par
In this work, the main objective is to predict the maneuver intention of a subject vehicle at each time frame. The intention can be lane keeping, left lane change and right lane change at maneuver level.

\subsection{Problem definition}
In this work the vehicle whose lane-change intention is predicted is referred to as the \emph{subject} or \emph{target} vehicle. The vehicles located in its neighborhood and affecting driving decisions are assumed to be the surrounding vehicles. According to the lane involvement of the subject vehicle, up to~8 surrounding vehicles are considered: the lead and following vehicles in the current lane as well as the lead, alongside and following vehicles in both left and right adjacent lanes.
\par
The lane-changing intention anticipation problem is formulated as a classification problem.
In each moment, the driver's intention is placed into one of three classes: lane keeping ($\mathrm{K}$), left lane change ($\mathrm{L}$), or right lane change ($\mathrm{R}$). 
\par
Let~$\O_t\in\mathbb{R}^n$ denote the observation vector at frame~$t\in\{1,\ldots,T\}$, where~$T$ denotes the final time of the considered track.
Here, $\O_t$ represents the available observation information about the subject vehicle and its surrounding vehicles.
The observation vector can include kinematic information of the subject vehicle, such as velocity and acceleration, and data on the interaction with the surrounding vehicles, such as relative distance and relative speed. Details on the considered observations, based on the available data in the~highD dataset, are given in the next section.
\par
We would like to predict for a given time horizon~$h\in\{1,\ldots,T-1\}$, whether a left change or a right change occurs within the next~$h$ time steps starting from the current frame~$t\in\{1,\ldots,T-h\}$. We model the driver intention as a discrete-state process~$\{I_t\}_{t=1}^T$ (i.e.~a sequence of discrete random variables on a probability space~$(\Omega,\mathcal{A},\mathbb{P})$) with~$I_t\in\{\mathrm{K},\mathrm{L},\mathrm{R}\}$ for all~$t\in\{1,\ldots,T-h\}$. Each~$\I_t$ describes the (unknown) true intention at frame~$t$. The value of~$I_t$ is interpreted as follows:
\begin{itemize}
    \item $I_t=\mathrm{K}$ if and only if \emph{no lane change} occurs within the frames~$t,t+1,\ldots,t+h$ ($\mathrm{K}$ refers to keeping the lane),
    \item $I_t=\mathrm{L}$ if and only if a \emph{lane change left} occurs within the frames~$t,t+1,\ldots,t+h$,
    \item $I_t=\mathrm{R}$ if and only if a \emph{lane change right} occurs within the frames~$t,t+1,\ldots,t+h$.
\end{itemize}
We would like to estimate for every~$t\in\{1,\ldots,T-h\}$ the distribution of~$I_t$ given the observation history~$\O_1,\ldots,\O_t$ up to the current frame:
    \begin{align*}
        \mathbb{P}\left[I_t=\mathrm{K}\mid\O_1,\ldots,\O_t\right] &= \mathbb{P}\left[\text{no lane change occurs within the next }h\text{ frames}\mid\O_1,\ldots,\O_t\right], \\
        \mathbb{P}\left[I_t=\mathrm{L}\mid\O_1,\ldots,\O_t\right] &= \mathbb{P}\left[\text{a left change occurs within the next }h\text{ frames}\mid\O_1,\ldots,\O_t\right], \\
        \mathbb{P}\left[I_t=\mathrm{R}\mid\O_1,\ldots,\O_t\right] &= \mathbb{P}\left[\text{a right change occurs within the next }h\text{ frames}\mid\O_1,\ldots,\O_t\right].
    \end{align*}
The prediction therefore refers to a future time interval of length~$h$. We thus aim to predict which intention will occur within the next~$h$ seconds. In this work, prediction horizons of~$h \in \{1,2,3,4,5,6\}$ seconds are considered (at a frame rate of~25 frames per second in the~highD dataset).

\subsection{Available measurements and observations}\label{sec:available-measurements-and-observations}
In this work, we use the~highD dataset~\cite{Krajewski2018} which was introduced in~2018 and collected with a drone-based aerial recording setup.
It contains naturalistic vehicle trajectories that cover a total driven distance of approximately~45.000\,km.
The data was acquired on highway segments, during daytime and under sunny weather conditions.
Since the dataset provides detailed vehicle information, including position, velocity, acceleration, lane-related information and interaction-based information, it is well suited for lane-change intention anticipation tasks.
As mentioned above, the~highD dataset provides observations at a rate of~$25$ frames per second.
\par
We organized the feature set into three major categories for consistency, reproducibility, and scalability: the kinematic features of the subject vehicle, the interaction features with its surrounding vehicles, and the occupancy flags of the neighboring positions.
We describe each category below.
\newline
\newline
\textbf{Kinematic features}
\newline
The first category describes the kinematic features of the subject vehicle whose intention is predicted at each frame. It contains the longitudinal and lateral velocities and the longitudinal and lateral accelerations of the subject vehicle. The observation vector is defined as
\begin{align*}
    \O_t^{\mathrm{kin}} = \big[v_{x,t},\ v_{y,t},\ a_{x,t},\ a_{y,t}\big] \in \mathbb{R}^{4}.
\end{align*}
\textbf{Surrounding-vehicle features}
\newline
The surrounding vehicle information group has a modular structure: each neighboring position is described by its own set of features, and these sets are kept separate. We first consider eight possible neighbor positions, that is
\begin{align*}
    \mathcal{S} = \{P,\, F,\, LP,\, LF,\, RP,\, RF,\, RA,\, LA\}.
\end{align*}
Here, $P$ and~$F$ denote the preceding and following vehicles in the current lane. The terms~$LP$, $LF$, and~$LA$ denote the preceding, following, and alongside vehicles in the left lane, while $RP$, $RF$, and~$RA$ refer to the same positions in the right lane. This set is designed to represent the complete neighborhood configuration around the subject vehicle. Since the alongside positions ($LA$ and~$RA$) did not provide a meaningful contribution to the separation of the intention classes (see~\Cref{sec:feature-selection}) and to reduce the dimensionality of the observations and thus simplify the models, we remove the alongside vehicles and mainly work with the reduced set of six positions
\begin{align*}
    \mathcal{S}' = \{P,\, F,\, LP,\, LF,\, RP,\, RF\}.
\end{align*}
For each neighbor position~$j \in \mathcal{S}$, two features are available: the relative longitudinal distance~$\Delta x_t^{(j)}$ and the relative longitudinal velocity~$\Delta v_t^{(j)}$. If there is no vehicle in a neighbor position, fixed default values are assigned. The surrounding vehicle information is defined as
\begin{align*}
    \O_t^{\mathrm{full-sur}} = \big[{\Delta x_t^{(j)},\ \Delta v_t^{(j)}}\big]_{j\in\mathcal{S}} \in \mathbb{R}^{16}\qquad\text{and}\qquad \O_t^{\mathrm{red-sur}} = \big[{\Delta x_t^{(j)},\ \Delta v_t^{(j)}}\big]_{j\in\mathcal{S}'} \in \mathbb{R}^{12}
\end{align*}
for the full neighbor set and the reduced set, respectively.
\newline
\newline
\textbf{Occupancy flags}
\newline
Lastly, for each neighbor position~$j\in\mathcal{S}$ we define an occupancy flag denoted by~$\delta_t^{(j)}\in\{0,1\}$ which indicates whether a vehicle exists in that position at time~$t$: 
\begin{align*}
    \delta_t^{(j)} = \begin{cases}
    1, & \text{if a vehicle exists in position } j \text{ at time } t, \\
    0, & \text{otherwise}.
    \end{cases}
\end{align*}
The occupancy flag allows the model to distinguish a position that is truly empty from one whose vehicle is present but far away.
The vector of all occupancy flags is defined as
\begin{align*}
    \O_t^{\mathrm{occ}} = \big[\delta_t^{(j)}\big]_{j\in\mathcal{S}} \in \mathbb{R}^{8}.
\end{align*}
\newline
\newline
\textbf{Feature configuration}
\newline
By combining these three categories of features, four different ways to set up our data, each with more information than the last, are available (in brackets we mention the abbreviations of the four feature configurations used in our numerical experiments in~\Cref{sec:experiments}):
\begin{enumerate}
    \item Kinematic set (\emph{kinematics}): This feature set only includes the subject vehicle's~4 kinematic features~$\O_t^{\mathrm{kin}}$.
    \item Compact interaction set (\emph{compact}): This observation set includes the four kinematic features of the subject vehicle and the reduced set of six surrounding vehicles (including twelve features). Therefore, the compact interaction feature set contains~16 observations in total and is given by~$\O_t^{\mathrm{kin}}$ together with~$\O_t^{\mathrm{red-sur}}$.
    \item Full continuous interaction set (\emph{full-cont.}): This set uses the subject vehicle's kinematic features and the eight surrounding vehicles. This representation includes~20 features in total and collects~$\O_t^{\mathrm{kin}}$ as well as~$\O_t^{\mathrm{full-sur}}$.
    \item Full continuous interaction set with occupancy flags (\emph{full+flags}): This feature set combines the subject vehicle's kinematic features, eight surrounding vehicle features and the corresponding eight occupancy flags. The complete set thus consists of~28 features and uses~$\O_t^{\mathrm{kin}}$, $\O_t^{\mathrm{full-sur}}$ and~$\O_t^{\mathrm{occ}}$ as observations.
\end{enumerate}
This structure is chosen on purpose.
First, it simplifies the features by considering individual measurements as observations.
Second, and more importantly, it allows the models to handle missing data, because each neighbor position is a self-defined block, a missing surrounding observation can be either marginalized out or imputed.
In this way, the configuration of features supports the missing-data strategies.
The complete candidate feature set used in this study is summarized in~\Cref{tab:selected_features}.
\begin{table}[htbp]
    \centering
    \small
    \renewcommand{\arraystretch}{1.18}
    \begin{tabularx}{\textwidth}{@{}c l c X@{}}
        \toprule
        \textbf{ID} & \textbf{Symbol} & \textbf{Unit} & \textbf{Description of feature} \\
        \midrule
        
        \multicolumn{4}{@{}l}{\textbf{Subject vehicle features}} \\
        \addlinespace[2pt]
        1  & $v_x$ & m/s & Longitudinal velocity of the subject vehicle \\
        2  & $v_y$ & m/s & Lateral velocity of the subject vehicle \\
        3  & $a_x$ & m/s$^2$ & Longitudinal acceleration of the subject vehicle \\
        4  & $a_y$ & m/s$^2$ & Lateral acceleration of the subject vehicle \\
        
        \addlinespace[5pt]
        \multicolumn{4}{@{}l}{\textbf{Current lane neighbor features}} \\
        \addlinespace[2pt]
        5  & $\Delta x^{(P)}$ & m 
        & Longitudinal gap to the preceding vehicle in the current lane \\
        6  & $\Delta v^{(P)}$ & m/s 
        & Relative longitudinal speed to the preceding vehicle in the current lane \\
        7  & $\delta^{(P)}$ & binary 
        & Occupancy of the preceding vehicle in the current lane \\
        8  & $\Delta x^{(F)}$ & m 
        & Longitudinal gap to the following vehicle in the current lane \\
        9  & $\Delta v^{(F)}$ & m/s 
        & Relative longitudinal speed to the following vehicle in the current lane \\
        10 & $\delta^{(F)}$ & binary 
        & Occupancy of the following vehicle in the current lane \\
        
        \addlinespace[5pt]
        \multicolumn{4}{@{}l}{\textbf{Left lane neighbor features}} \\
        \addlinespace[2pt]
        11  & $\Delta x^{(LP)}$ & m 
        & Longitudinal gap to the preceding vehicle in the left lane \\
        12 & $\Delta v^{(LP)}$ & m/s 
        & Relative longitudinal speed to the preceding vehicle in the left lane \\
        13 & $\delta^{(LP)}$ & binary 
        & Occupancy of the preceding vehicle in the left lane \\
        14 & $\Delta x^{(LF)}$ & m 
        & Longitudinal gap to the following vehicle in the left lane \\
        15 & $\Delta v^{(LF)}$ & m/s 
        & Relative longitudinal speed to the following vehicle in the left lane \\
        16 & $\delta^{(LF)}$ & binary 
        & Occupancy of the following vehicle in the left lane \\
        17 & $\Delta x^{(LA)}$ & m 
        & Longitudinal gap to the alongside vehicle in the left lane \\
        18 & $\Delta v^{(LA)}$ & m/s 
        & Relative longitudinal speed to the alongside vehicle in the left lane \\
        19 & $\delta^{(LA)}$ & binary 
        & Occupancy of the alongside vehicle in the left lane \\
        
        \addlinespace[5pt]
        \multicolumn{4}{@{}l}{\textbf{Right lane neighbor features}} \\
        \addlinespace[2pt]
        20 & $\Delta x^{(RP)}$ & m 
        & Longitudinal gap to the preceding vehicle in the right lane \\
        21 & $\Delta v^{(RP)}$ & m/s 
        & Relative longitudinal speed to the preceding vehicle in the right lane \\
        22 & $\delta^{(RP)}$ & binary 
        & Occupancy of the preceding vehicle in the right lane \\
        23 & $\Delta x^{(RF)}$ & m 
        & Longitudinal gap to the following vehicle in the right lane \\
        24 & $\Delta v^{(RF)}$ & m/s 
        & Relative longitudinal speed to the following vehicle in the right lane \\
        25 & $\delta^{(RF)}$ & binary 
        & Occupancy of the following vehicle in the right lane \\
        26 & $\Delta x^{(RA)}$ & m 
        & Longitudinal gap to the alongside vehicle in the right lane \\
        27 & $\Delta v^{(RA)}$ & m/s 
        & Relative longitudinal speed to the alongside vehicle in the right lane \\
        28 & $\delta^{(RA)}$ & binary 
        & Occupancy of the alongside vehicle in the right lane \\
        \bottomrule
    \end{tabularx}
    \caption{Complete candidate features in the highD dataset~\cite{Krajewski2018}.}
    \label{tab:selected_features}
\end{table}
\par
\begin{figure}[htbp]
  \centering
  \begin{subfigure}[t]{0.46\linewidth}\centering
    \resizebox{\linewidth}{!}{%
    \begin{tikzpicture}[
        >=Latex,
        veh/.style   ={rounded corners=2pt, draw=black!65, minimum width=1.5cm,
                       minimum height=0.7cm, inner sep=1pt},
        ego/.style   ={veh, draw=blue!70!black, line width=1pt, fill=blue!12, font=\small},
        along/.style ={veh, fill=black!22},          
        nb/.style    ={veh, fill=white},
        lane/.style  ={black!40, dash pattern=on 6pt off 5pt, line width=0.7pt},
        bound/.style ={black!75, line width=1.1pt},
      ]
      \def\xmin{-5.0}\def\xmax{5.0}
      \def\yL{1.3}\def\yR{-1.3}
      \def\xP{3.0}                                  
      \useasboundingbox (\xmin,-2.95) rectangle (\xmax,2.7);   

      \fill[black!3] (\xmin,-1.95) rectangle (\xmax,1.95);
      \draw[bound] (\xmin,1.95) -- (\xmax,1.95);
      \draw[bound] (\xmin,-1.95) -- (\xmax,-1.95);
      \draw[lane] (\xmin,0.65) -- (\xmax,0.65);
      \draw[lane] (\xmin,-0.65) -- (\xmax,-0.65);
      \draw[->,black!70,line width=0.9pt] (\xmin+0.25,2.35) -- ++(1.4,0)
            node[right,font=\scriptsize,black!70]{driving direction};

      \node[nb]    at (\xP,0)    {$P$};
      \node[ego]   at (0,0)      {subject};
      \node[nb]    at (-2.6,0)   {$F$};
      \node[nb]    at (2.2,\yL)  {$LP$};
      \node[along] at (0.5,\yL)  {$LA$};
      \node[nb]    at (-3.3,\yL) {$LF$};
      \node[nb]    at (3.7,\yR)  {$RP$};
      \node[along] at (-0.5,\yR) {$RA$};
      \node[nb]    at (-2.3,\yR) {$RF$};

      \draw[->,black!75,line width=0.9pt] (\xP+0.9,0) -- (\xP+1.75,0);
      \node[above=-1pt,font=\scriptsize,black!75] at ({\xP+1.325},0.02) {$\Delta v_t^{(P)}$};
      \draw[black!45,dotted,line width=0.6pt] (0,-1.95) -- (0,-2.75);
      \draw[black!45,dotted,line width=0.6pt] (\xP,-1.95) -- (\xP,-2.75);
      \draw[<->,black!75,line width=0.8pt] (0,-2.6) -- (\xP,-2.6);
      \node[below,font=\scriptsize,black!75] at ({\xP/2},-2.6) {$\Delta x_t^{(P)}$};
    \end{tikzpicture}}
    \caption{full surrounding}
    \label{fig:surround-full}
  \end{subfigure}\hfill
  \begin{subfigure}[t]{0.46\linewidth}\centering
    \resizebox{\linewidth}{!}{%
    \begin{tikzpicture}[
        >=Latex,
        veh/.style   ={rounded corners=2pt, draw=black!65, minimum width=1.5cm,
                       minimum height=0.7cm, inner sep=1pt},
        ego/.style   ={veh, draw=blue!70!black, line width=1pt, fill=blue!12, font=\small},
        along/.style ={veh, fill=black!22},
        nb/.style    ={veh, fill=white},
        ghost/.style ={veh, draw=black!40, densely dashed, fill=black!2,
                       text=black!55, font=\scriptsize},   
        lane/.style  ={black!40, dash pattern=on 6pt off 5pt, line width=0.7pt},
        bound/.style ={black!75, line width=1.1pt},
      ]
      \def\xmin{-5.0}\def\xmax{5.0}
      \def\yL{1.3}\def\yR{-1.3}
      \useasboundingbox (\xmin,-2.95) rectangle (\xmax,2.7);   

      \fill[black!3] (\xmin,-1.95) rectangle (\xmax,1.95);
      \draw[bound] (\xmin,1.95) -- (\xmax,1.95);
      \draw[bound] (\xmin,-1.95) -- (\xmax,-1.95);
      \draw[lane] (\xmin,0.65) -- (\xmax,0.65);
      \draw[lane] (\xmin,-0.65) -- (\xmax,-0.65);
      \draw[->,black!70,line width=0.9pt] (\xmin+0.25,2.35) -- ++(1.4,0)
            node[right,font=\scriptsize,black!70]{driving direction};

      \node[ghost] at (2.4,0)    {$\delta_t^{(P)}{=}0$};
      \node[ego]   at (0,0)      {subject};
      \node[nb]    at (-3.2,0)   {$F$};
      \node[nb]    at (3.4,\yL)  {$LP$};
      \node[along] at (-0.4,\yL) {$LA$};
      \node[ghost] at (-2.7,\yL) {$\delta_t^{(LF)}{=}0$};
      \node[nb]    at (2.6,\yR)  {$RP$};
      \node[ghost] at (0.6,\yR)  {$\delta_t^{(RA)}{=}0$};
      \node[nb]    at (-3.4,\yR) {$RF$};
    \end{tikzpicture}}
    \caption{partial observation}
    \label{fig:surround-partial}
  \end{subfigure}
  \caption{Top-view schematic of the neighborhood around a subject vehicle.
  (a) all eight neighbors present; (b) under partial observation, an absent
  neighbor~$j\in\mathcal{S}$ carries the occupancy flag~$\delta_t^{(j)}=0$.}
  \label{fig:surround}
\end{figure}
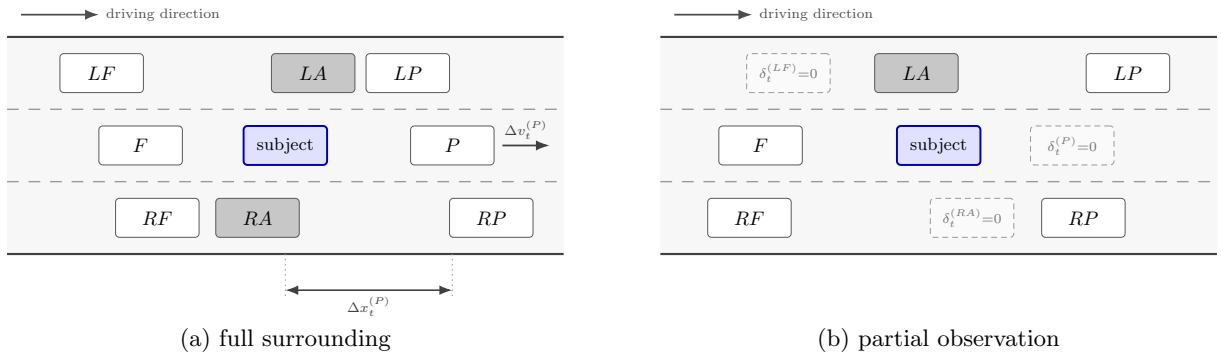

We further provide an overview of the surrounding of a vehicle in~\Cref{fig:surround}, including a sketch of a full surrounding (left part of the figure) and the more realistic scenario of partial observations (right part of the figure).
The blue box indicates the subject vehicle itself (containing the four kinematic features).
The white boxes represent the neighboring vehicles taken into account for the compact interaction set.
Additionally, the gray boxes refer to the alongside vehicles which are added to the compact set in order to obtain the full continuous interaction set.
Further including the occupancy flags (indicated in~\Cref{fig:surround-partial}) yields the entire feature set.

\subsection{Model for observation dropout}\label{sec:model-observation-dropout}
In order to make the data more realistic, we assume that not all observations are available at any time.
In practice we expect to face missing information regularly and the driver intention prediction should be robust against such corrupted data.
We thus consider the following strategy for simulating randomly dropped observations:
First of all, we (mainly) assume that the kinematic features~$\O_t^\mathrm{kin}$ of the subject vehicle are always present and never dropped (in~\Cref{sec:temporal-consistency} we also consider a full blackout in one of our experiments).
It is important to note that the kinematic features are the ones having the largest influence on the prediction of the driver intention (this will be investigated in detail in~\Cref{sec:experiments}).
Since we expect those observations to be the most stable and reliable features, we do not drop them here in the majority of our studies.
The surround, in contrast, might be subject to occlusion or other sensor disturbances, which we model by dropout of certain parts of the associated observations.
To be more precise, we consider the different neighbor slots separately and assume that if a slot is dropped for a frame, then all of its associated features are missing in that observation vector.
We expect this to be a more realistic model for observation damage than individual missing features.
\par
The random process for generating the corrupted data generates a mask for every frame and every slot, indicating whether the observations of the respective slot are missing in the considered frame.
For the actual dropout per frame and slot we consider two different scenarios:
On the one hand, each slot per frame is dropped independently with probability~$\rho\in[0,1]$.
In this model, no temporal structure in the dropout is present and slots drop and reappear on a frame by frame basis.
We refer to this dropout type as~\emph{i.i.d.~dropout}.
A probably more realistic corruption model includes also a temporal correlation of the dropout within each slot.
The motivation for this model is the assumption that a neighboring vehicle's information is typically missing over a longer period of time instead of a single frame.
In order to model this behavior, a Markov chain with two states (``present'' and ``dropped'') is considered for each slot.
A mean burst length of~$b$ frames then leads to the transition probabilities~$p_1=1-1/b$ as the probability for a slot to remain dropped once it was dropped and~$p_2=\rho\cdot(1-p_1)/(1-\rho)$ as the probability of transitioning to the dropped state.
We therefore obtain the following transition matrix for the Markov chain:
\begin{align*}
    \begin{bmatrix}
        1-\frac{\rho}{(1-\rho)b} & \frac{\rho}{(1-\rho)b} \\
        \frac{1}{b} & 1-\frac{1}{b}
    \end{bmatrix}.
\end{align*}
The stationary distribution of this Markov chain can be computed by determining the left eigenvector of the transition matrix for the eigenvalue~$1$ whose sum is equal to~$1$.
This vector is then given as
\begin{align*}
    \begin{bmatrix}
        1-\rho \\
        \rho
    \end{bmatrix},
\end{align*}
which means that the stationary distribution is chosen in such a way that the stationary drop-probability is exactly equal to~$\rho$ -- similar to the case of~i.i.d.~dropout discussed before.
In our numerical experiments in~\Cref{sec:experiments} we restrict ourselves to a mean burst length of~$b=1$\,s.
This dropout mechanism will be called~\emph{bursty dropout} in the following.
\par
The approach based on a Gaussian mixture model, see~\Cref{sec:GMM}, will see the entire, uncorrupted data during training since it handles dropout naturally via marginalization.
The corruption process described above is applied to it only at inference.
In contrast, the discriminative methods in~\Cref{sec:scaled-likelihood} are additionally trained on dropout-augmented data:
To this end, each training instant is augmented with additional copies whose slots are dropped at a random rate~$\rho\sim\mathcal{U}[0,\rho_{\max}]$ and imputed with the same rule used at inference.
As these are per-frame classifiers, the augmentation spans a range of dropout severities but no temporal structure.
Bursty dropout cannot be simulated during training and therefore appears only during inference.

\section{Recursive Bayesian filtering for long-term lane-change intention anticipation}\label{sec:recursive-bayesian-filtering}
In this section we outline the basic ideas and components of our proposed driver intention prediction model.
We first introduce the recursive Bayesian filtering approach that provides a probabilistic model for driver intention.
Afterwards, we discuss practical issues such as high-dimensional observation spaces, different emission models, handling of missing information due to sensor dropout and calibration of the resulting probabilities.

\subsection{Description of the methodology}
For the driver intention prediction we would like to estimate the probability distribution of the intention~$\I_t$ at frame~$t$ based on the previous and current observations~$\O_{t-w},\ldots,\O_t$ for a window length~$w\geq 0$.
We are therefore interested in~$\P{\I_t\mid\O_{t-w},\ldots,\O_t}$.
This quantity is, however, not directly accessible and training a machine learning model based on an entire history that also grows over time becomes complicated (transformers with their attention mechanism try to achieve this, but we focus on a different approach here).
Hence, we would like to rewrite~$\P{\I_t\mid\O_{t-w},\ldots,\O_t}$ in such a way that we obtain a formula involving~$\P{\I_{t-1}\mid\O_{t-w},\ldots,\O_{t-1}}$.
We can then insert the same formula recursively.
\par
The derivation is based on the following two assumptions:
\begin{align}\label{equ:assumptions}
    \begin{split}
        \P{\O_t\mid\I_t,\O_{t-w},\ldots,\O_{t-1}} &= \P{\O_t\mid\I_t}, \\
        \P{\I_t\mid\I_{t-1},\O_{t-w},\ldots,\O_{t-1}} &= \P{\I_t\mid\I_{t-1}}.
    \end{split}
\end{align}
The first assumption implies that the observations depend only on the current intention.
The second assumption is a first order Markov property for the intention, i.e.~the current intention depends only on the previous intention and not the entire history of previous observations.
\begin{remark}
    The assumptions in~\eqref{equ:assumptions} will hold most likely only \emph{approximately} in practice, because the intention is not instantaneous but considered over the subsequent~$h$ frames.
    Therefore, the observation at time~$t$ will usually depend on the previous observations since those had an influence on previous intentions leading to current and future maneuvers.
    We nevertheless state these assumptions as modeling basis for our approach and assume that they hold up to negligible deviations.
\end{remark}
\noindent With these assumptions at hand and the definition~$Z_t=\P{\O_t\mid\O_{t-w},\ldots,\O_{t-1}}$, we can derive
\begin{align*}
    \P{\I_t\mid\O_{t-w},\ldots,\O_t} &= \P{\I_t\mid\O_t,\O_{t-w},\ldots,\O_{t-1}} \\
    &= \frac{\P{\O_t\mid\I_t,\O_{t-w},\ldots,\O_{t-1}}\cdot\P{\I_t\mid\O_{t-w},\ldots,\O_{t-1}}}{\P{\O_t\mid\O_{t-w},\ldots,\O_{t-1}}} \\
    &= \frac{\P{\O_t\mid\I_t}\cdot\P{\I_t\mid\O_{t-w},\ldots,\O_{t-1}}}{\P{\O_t\mid\O_{t-w},\ldots,\O_{t-1}}} \\
    &= \frac{1}{Z_t}\P{\O_t\mid\I_t}\cdot\P{\I_t\mid\O_{t-w},\ldots,\O_{t-1}} \\
    &= \frac{1}{Z_t}\P{\O_t\mid\I_t}\cdot\sum\limits_{\I_{t-1}\in\{\mathrm{K},\mathrm{L},\mathrm{R}\}}\P{\I_t\mid\I_{t-1},\O_{t-w},\ldots,\O_{t-1}}\cdot\P{\I_{t-1}\mid\O_{t-w},\ldots,\O_{t-1}} \\
    &= \frac{1}{Z_t}\P{\O_t\mid\I_t}\cdot\sum\limits_{\I_{t-1}\in\{\mathrm{K},\mathrm{L},\mathrm{R}\}}\P{\I_t\mid\I_{t-1}}\cdot\P{\I_{t-1}\mid\O_{t-w},\ldots,\O_{t-1}},
\end{align*}
where we reordered the observations in the first step, applied Bayes' Theorem in the second step, used the first assumption from~\eqref{equ:assumptions} in the third step, the definition of~$Z_t$ in the fourth step, the law of total probability in the fifth step, and finally inserted the second assumption from~\eqref{equ:assumptions} in the last step.
The recursive Bayesian update rule is therefore given as
\begin{align}\label{equ:recursive-bayesian-update}
    \P{\I_t \mid \O_{t-w},\ldots,\O_t} = \frac{1}{Z_t}\P{\O_t \mid \I_t}\cdot\sum\limits_{\I_{t-1}\in\{\mathrm{K},\mathrm{L},\mathrm{R}\}}\P{\I_t \mid \I_{t-1}}\cdot \P{\I_{t-1} \mid \O_{t-w},\ldots,\O_{t-1}},
\end{align}
where~$Z_t=\P{\O_t\mid\O_{t-w},\ldots,\O_{t-1}}$ is a normalization constant.
The transition probability~$\P{\I_t\mid\I_{t-1}}$ is determined empirically based on the dataset.
In practice, the previous posterior~$\P{\I_{t-1} \mid \O_{t-w},\ldots,\O_{t-1}}$ can now be replace by~\eqref{equ:recursive-bayesian-update} after shifting the time index by one (and reducing the window length~$w$ by one as well).
This can be done recursively until reaching the frame~$t-w$ and the recursion finishes.
As the base case, i.e.~the prior at frame~$t-w$, we choose the uniform distribution over the intention classes.
The initial prior is therefore uninformative by construction and the entire recursion is determined by the transition probabilities and the observations.
\par
Whenever classification of the intention is considered, we take the class with the highest probability and return that one as prediction.

\subsection{Interpretation as a hidden Markov model}
The recursive Bayesian filtering model introduced in the previous section can also be interpreted as a hidden Markov model~\cite{Li2023}, where the intentions correspond to the hidden states, the observations (also referred to as~\emph{emissions}) are generated according to the hidden state, and the emission model is the likelihood term that connects the hidden state to the observations.
We discuss the connection in detail in this section.
\par
In this context, the recursive Bayesian filtering can be represented by the hidden Markov model parameters~$\lambda = (A,B,\pi)$, where~$A$ denotes the state transition probability matrix, $B$ is the observation (\emph{emissions}) model, and~$\pi$ is the initial state distribution.
These components of the model are defined as follows:

\begin{enumerate}
    \item Hidden state set: A finite set~$S = \{s_1,s_2,\dots, s_N\}$ of~$N$ possible hidden states.
    At every time step~$t$, the model is in one of these states, denoted as~$I_t \in S$.
    In the driver intention prediction application, we have~$S=\{\mathrm{K},\mathrm{L},\mathrm{R}\}$ and therefore~$N=3$.
    \item Observation space: Let~$\O\subseteq \mathbb{R}^{n}$ denote the observation space, where~$n$ is the dimension of the observation vector.
    At every time step~$t$, an observation~$\O_t\in\O$ is emitted by the model.
    \item State transition probability matrix: The temporal evolution of the hidden driver intention is represented by the state transition probability matrix~$A = [a_{ij}] \in [0,1]^{N\times N}$, where
    \begin{align*}
        a_{ij} = \P{I_t=s_j \mid I_{t-1}=s_i}\qquad\text{for }1 \leq i,j \leq N.
    \end{align*}
    Here, $a_{ij}$ denotes the probability of transitioning from hidden state~$s_i$ at time~$t-1$ to hidden state~$s_j$ at time~$t$.
    Since each row of the transition matrix represents a probability distribution, we have~$a_{i,j}\geq0$ for all~$i,j=1,\ldots,N$ and~$\sum_{j=1}^{N} a_{i,j}=1$ for all~$i=1, \dots,N$.
    For~$I_t=\{\mathrm{K},\mathrm{L},\mathrm{R}\}$ the transition matrix takes the form
    \begin{align*}
        A = \begin{bmatrix}
            a_{KK} & a_{KL} & a_{KR} \\
            a_{LK} & a_{LL} & a_{LR} \\
            a_{RK} & a_{RL} & a_{RR} 
        \end{bmatrix}\in[0,1]^3.
    \end{align*}
    For instance, $a_{KL} = \P{I_t=L \mid I_{t-1}=K}$ denotes the probability that the hidden state changes from lane keeping to a left-lane change intention between two frames.
    This structure represents the temporal persistence of driving intentions and reduces unrealistic frame-to-frame changes in the anticipated intention.
    The entries of the state transition matrix are derived from data.
    \item Emission model: For each hidden state~$s_j \in S$, the emission model defines the state-conditional density of the current observation vector.
    It is denoted as
    \begin{align*}
        b_j(\O_t) = \P{\O_t \mid I_t = s_j}\qquad\text{for }j=1,\dots, N.
    \end{align*}
    The quantity~$b_j(\O_t)$ represents the likelihood of obtaining the observation vector~$\O_t$ when the hidden intention is~$s_j$.
    Two different methods for modeling this likelihood will be introduced in the following section.
    \item Initial state distribution: The initial state distribution specifies the prior probabilities of the hidden driver intention before the first observation is incorporated. It is given as~$\pi = [\pi_1, \pi_2, \dots, \pi_N]$, where~$\pi_i = \P{I_{t-w} = s_i}$ for all~$i=1,\dots, N$.
    The initial probabilities satisfy~$\pi_i\geq 0$ for all~$i=1,\ldots,N$ and~$\sum_{i=1}^{N}\pi_i=1$.
    As mentioned above, a uniform distribution is chosen as the initial state distribution in our application and we therefore have~$\pi = [\pi_\mathrm{K},\pi_\mathrm{L},\pi_\mathrm{R}]=[1/3,1/3,1/3]$.
\end{enumerate}

\noindent In the recursive Bayesian update formula from~\eqref{equ:recursive-bayesian-update}, the terms from the hidden Markov model can be identified as
\begin{align*}
    \P{\I_t \mid \O_{t-w},\ldots,\O_t} = \frac{1}{Z_t}\underbrace{\P{\O_t \mid \I_t}}_{\text{emission model}}\cdot\sum\limits_{\I_{t-1}\in\{\mathrm{K},\mathrm{L},\mathrm{R}\}}\underbrace{\P{\I_t \mid \I_{t-1}}}_{\text{transition matrix}}\cdot \underbrace{\P{\I_{t-1} \mid \O_{t-w},\ldots,\O_{t-1}}}_{\text{recursion}}.
\end{align*}
\Cref{fig:filter} visualizes the hidden Markov model as a posterior distribution of the intention classes, evolving over time, based on continuously collected observations and the (time- and observation-independent) transition matrix.
In this example, the posterior over~$\I_t$ concentrates on the intention class~$\mathrm{L}$, indicating a change to the left lane.
Starting from a uniform prior distribution at time~$t-w$, the model accumulates evidence for class~$\mathrm{L}$ over a window of length~$w$, taking the incoming observations as well as the previous posterior distribution into account.
\definecolor{colK}{RGB}{140,140,140}   
\definecolor{colL}{RGB}{35,110,190}    
\definecolor{colR}{RGB}{225,140,40}    
\providecommand{\bel}[5]{%
  \begin{scope}[shift={(#1,0)}]
    \draw[rounded corners=2pt, #5, fill=black!3] (-0.82,-0.05) rectangle (0.82,1.12);
    \fill[colK] (-0.54,0) rectangle (-0.26,#2);
    \fill[colL] (-0.14,0) rectangle ( 0.14,#3);
    \fill[colR] ( 0.26,0) rectangle ( 0.54,#4);
    \draw[black!35,line width=0.4pt] (-0.68,0) -- (0.68,0);
    \node[font=\tiny,text=black!65,below=-1pt] at (-0.40,0) {K};
    \node[font=\tiny,text=black!65,below=-1pt] at ( 0.00,0) {L};
    \node[font=\tiny,text=black!65,below=-1pt] at ( 0.40,0) {R};
  \end{scope}%
}
\begin{figure}[htbp]
  \centering
  \resizebox{\linewidth}{!}{%
  \begin{tikzpicture}[
      >=Latex,
      obs/.style={rounded corners=2pt, draw=black!55, fill=white, minimum width=1.15cm,
                  minimum height=0.58cm, inner sep=1pt, font=\small},
    ]
    \def\xa{0}\def\xb{3.0}\def\xc{6.8}\def\xd{9.8}

    \node[font=\scriptsize,text=black!70,anchor=east] at (-1.4,2.55) {observation $\mathcal{O}_t$};
    \node[font=\scriptsize,text=black!70,anchor=east] at (-1.4,0.55) {posterior over $I_t$};

    \node[obs] at (\xb,2.55) {$\mathcal{O}_{t-w+1}$};
    \node[obs] at (\xc,2.55) {$\mathcal{O}_{t-1}$};
    \node[obs] at (\xd,2.55) {$\mathcal{O}_{t}$};

    \foreach \x in {\xb,\xc,\xd}{
      \draw[->,black!70,line width=0.8pt] (\x,2.22) -- (\x,1.20);
    }

    \bel{\xa}{0.333}{0.333}{0.333}{black!18}   
    \bel{\xb}{0.42}{0.48}{0.10}{black!18}
    \bel{\xc}{0.22}{0.72}{0.06}{black!18}
    \bel{\xd}{0.10}{0.86}{0.04}{black!18}

    \draw[->,black!70,line width=0.8pt] (\xa+0.9,0.5) -- (\xb-0.9,0.5);
    \draw[->,black!70,line width=0.8pt] (\xc+0.9,0.5) -- (\xd-0.9,0.5);
    \draw[->,black!70,line width=0.8pt] (\xb+0.9,0.5) -- ({(\xb+\xc)/2-0.5},0.5);
    \node[font=\small,text=black!55] at ({(\xb+\xc)/2},0.48) {$\cdots$};
    \draw[->,black!70,line width=0.8pt] ({(\xb+\xc)/2+0.5},0.5) -- (\xc-0.9,0.5);
    \node[font=\scriptsize,text=black!70] at ({(\xa+\xb)/2},0.72) {$A$};
    \node[font=\scriptsize,text=black!70] at ({(\xc+\xd)/2},0.72) {$A$};

    \node[font=\scriptsize,text=black!60] at (\xa,-0.62) {$t\!-\!w$};
    \node[font=\scriptsize,text=black!60] at (\xb,-0.62) {$t\!-\!w\!+\!1$};
    \node[font=\scriptsize,text=black!60] at (\xc,-0.62) {$t\!-\!1$};
    \node[font=\scriptsize,text=black!60] at (\xd,-0.62) {$t$};

    \draw[->,black!45,line width=0.7pt] (-1.15,-1.05) -- (\xd+1.9,-1.05)
         node[right,font=\scriptsize,text=black!60]{time};
  \end{tikzpicture}}
  \caption{Sketch of the posterior evolution in the recursive Bayesian filtering approach for a left-change example.}
  \label{fig:filter}
\end{figure}
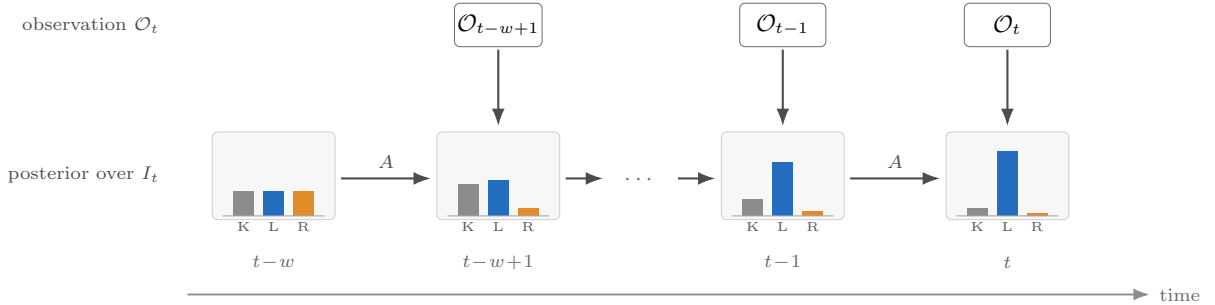

\subsection{Emission model}
In this section we describe two different emission models.
The first approach applies Gaussian mixture models to represent the likelihood term.
As we will see in our experiments in~\Cref{sec:experiments}, this method is limited to low-dimensional observations and struggles when dealing with many of the~28 features provided in the~highD dataset.
To circumvent this issue, the second strategy presented below models the emission distribution via an approximation of the likelihood by a scaled posterior term.

\subsubsection{Gaussian mixture models for the likelihood term}\label{sec:GMM}
A natural choice of the likelihood term~$\P{\O_t\mid\I_t}$ are Gaussian mixture models~(GMMs)~\cite{reynolds2009gaussian}.
Here, we consider a separate~GMM per intention class, that is,
\begin{align*}
    \P{\O_t\mid\I_t=s} = \sum\limits_{i=1}^{N_\mathrm{comp}} w_i\cdot\mathcal{N}(\O_t;\mu_i,\Sigma_i)\qquad\text{for }s\in\{\mathrm{K},\mathrm{L},\mathrm{R}\},
\end{align*}
where~$w_1,\ldots,w_{N_\mathrm{comp}}\geq 0$ are the mixture weights with~$w_1+\cdots+w_{N_\mathrm{comp}}=1$, $\mu_1,\ldots,\mu_{N_\mathrm{comp}}\in\R^n$ are the component means and~$\Sigma_1,\ldots,\Sigma_{N_\mathrm{comp}}\in\R^{n\times n}$ are the component covariances with~$\Sigma_i$ being a symmetric and positive definite matrix for all~$i=1,\ldots,{N_\mathrm{comp}}$.
The mixture weights as well as means and covariances per component are optimized using the expectation maximization~(EM) algorithm~\cite{dempster1977maximum}, which computes (approximately) the maximum likelihood estimate of the parameters given a training dataset.
\begin{remark}[GMMs in high dimensions]\label{rem:GMMs-high-dimensions}
    It is well-known that~GMMs might have difficulties in representing probability densities in high dimensions.
    Hence, when dealing with high-dimensional observation spaces, GMMs lead to issues with the conditioning of the~EM algorithm.
    As discussed in~\Cref{sec:available-measurements-and-observations}, there are different feature configurations of interest available in the~highD dataset.
    In~\Cref{sec:feature-selection} we therefore investigate the performance of the methods discussed in this section depending on the amount of features used as observations.
\end{remark}

\subsubsection{Scaled-likelihood trick and discriminative posteriors}\label{sec:scaled-likelihood}
As an alternative to representing the likelihood via a generative model, one can instead apply the~\emph{scaled-likelihood trick}, see for instance~\cite{Renals1994}, which uses the identity
\begin{align}\label{equ:scaled-likelihood-trick}
    \P{\O_t\mid\I_t} \propto \frac{\P{\I_t\mid\O_t}}{\P{\I_t}}
\end{align}
to turn a posterior probability from a discriminative classifier into a (scaled) likelihood.
In our numerical experiments, we approximate the posterior~$\P{\I_t\mid\O_t}$ by a histogram-based gradient boosting machine~(GBM)~\cite{ke2017lightgbm}, based on classification trees, as a discriminative posterior classifier.
The prior~$\P{\I_t}$ is the class prior of the (balanced) training set, i.e.~the sampling distribution under which the~GBM is trained, see below.
\par
Since the Bayesian update rule in~\eqref{equ:recursive-bayesian-update} requires normalization over the intention classes, the (unknown) common factor~$\P{\O_t}$, which is independent of~$\I_t$ and corresponds to the missing scaling in~\eqref{equ:scaled-likelihood-trick}, is not required here.
Therefore, the discriminative posterior can be chosen quite flexibly and we are not restricted to~GBMs as used in this work.
It is in particular not necessary to ensure that the approximation of~$\P{\O_t\mid\I_t}$ according to~\eqref{equ:scaled-likelihood-trick} is, given~$\I_t$, a probability density with respect to~$\O_t$.
\par
The main advantage of this approach is that it is applicable also to higher-dimensional observation vectors~$\O_t$ since a discriminative classifier such as the~GBM can handle those observations as input.
Moreover, the number of intention classes is limited to three in our application, allowing for efficient training of a respective~GBM as described below.
\par
For training of the~GBM, we generate a dataset of pairs~$(o_i,s_i)$, where~$o_i\in\mathbb{R}^n$ denotes the~$i$-th observation vector and~$s_i\in\{\mathrm{K},\mathrm{L},\mathrm{R}\}$ the corresponding intention in the upcoming~$h$ frames.
The training set is chosen as a balanced subset of the full highD dataset, which contains one data point per maneuver and an equal number of data points labeled as lane-keeping instances.
The~GBM classifier defines a conditional distribution over the intention classes~$\{\mathrm{K},\mathrm{L},\mathrm{R}\}$ via a softmax over its per-class boosting scores, that is
\begin{align*}
    \hat{\mathbb{P}}(s\mid o) = \mathrm{softmax}_s f(o) \coloneqq \frac{\exp(f_s(o))}{\sum\limits_{c\in\{\mathrm{K},\mathrm{L},\mathrm{R}\}}\exp(f_c(o))}.
\end{align*}
Here, $f = (f_K,f_L,f_R)\colon\mathbb{R}^n \to \mathbb{R}^3$ is the collection of three separate ensembles of trees, i.e.
\begin{align*}
    f_s(o) = \sum\limits_{m=1}^{M} T_{m,s}(o),
\end{align*}
where~$T_{m,s}\colon\mathbb{R}^n\to\mathbb{R}$ is the~$m$-th regression tree for class~$s$ and~$M\in\mathbb{N}$ denotes the number of trees.
The softmax is used in order to turn the raw score vector, generated by the additive tree ensembles~$f_s$, into a normalized probability distribution.
The model is trained by minimizing the multinomial cross-entropy~$-\sum_i \log \hat{\mathbb{P}}(s_i\mid o_i)$ on the training set, which is equivalent to maximizing the conditional likelihood of the normalized posterior.

\subsection{Dealing with missing observations in the different models}
In what follows, we consider several approaches to deal with dropout in the observation data.
Depending on the model used (GMM or~GBM with the scaled-likelihood trick), different strategies are reasonable.
\par
For the~GMM, it is possible to marginalize out analytically the components that are missing in the data.
This is the most natural way to handle dropout and essentially corresponds to the assumption that indeed no information is available about the missing features.
The~GMM with marginalization in particular does not impose any assumptions on the corruption model, which might be important in case the dropout characteristics are unknown during training and can therefore not be applied to the training data.
This marginalization is impossible when using the scaled-likelihood trick and a discriminative classifier.
Hence, we only pursue this approach for the~GMM as model of the likelihood.
\par
In contrast, one can also fill the missing observations by the default value that is used when a neighboring vehicle is absent in the original, uncorrupted dataset.
In the~highD dataset used in this study, a free neighbor slot is indicated by a distance to the respective (non-existent) neighboring car of~$\pm250$m, a velocity difference of~$0$ and the binary flag that the neighbor is present set to~$0$.
For both, the~GMM and the~GBM, one can fill the missing data based on these default values for absent neighboring vehicles.
We refer to this way of imputing observations as~``sentinel'' in the sense that the value signals a missing feature.
However, a major drawback in this case is that ``missing information'' is essentially treated similarly to actually missing neighboring vehicles -- the models cannot distinguish these two cases.
When marginalizing out the missing components, it is explicitly taken into account that the respective parts of the observations are not available.
\par
In addition to the marginalization and the filling with default values, one might also consider other methods to impute the dropout intervals.
For instance, in our numerical experiments in~\Cref{sec:experiments}, we compare the previously mentioned approaches to filling with mean values and to the \emph{Last Observation Carried Forward~(LOCF)} technique~\cite{Overall2009}.
The mean observation is here computed per feature on the frames of the training set.
In~LOCF, the last available observation is used to fill gaps in the data and is carried across all frames that suffer from dropout.

\subsection{Calibration of probabilities}
Probability calibration is a vital step in classification tasks, ensuring that predicted probabilities accurately reflect the true likelihood of an event~\cite{guo2017calibration}.
A well-calibrated model, when predicting an event with a probability of, for instance, 80\%, implies that approximately~80\% of samples assigned this probability should indeed belong to the predicted class~\cite{dawid1982well}.
However, if this condition is not satisfied, the model may implicitly lead practitioners or users to place unwarranted trust in miscalibrated predictions, potentially resulting in serious consequences.
The calibration of predictive models is particularly relevant in high-stakes scenarios such as autonomous driving, making it of the utmost importance to have a model that is both accurate and reliable (i.e., well-calibrated).
Choosing a simpler, yet more interpretable, model such as a GMM instead of a neural network with millions of parameters, could partially mitigate this miscalibration problem, since one of the main causes is precisely the large capacity of modern networks and their tendency to overfit~\cite{guo2017calibration}.
Nonetheless, we attempt to further improve the calibration of our approaches considered in this work and study the effect of different calibration methods.
\par
Calibration methods aim to transform raw classifier scores into more reliable probability estimates.
They can be broadly divided into three families: \textit{post-hoc calibration}, \textit{train-time calibration}, and \textit{uncertainty estimation methods}~\cite{wang2023calibrating}.
The first class, {post-hoc calibration} \cite{zadrozny2001obtaining,zadrozny2002transforming,kull2017beta,platt1999probabilistic}, operates after the model has been trained.
These methods leave the model parameters unchanged and instead use a held-out calibration dataset to adjust the output scores, thereby improving the alignment between predicted probabilities and observed frequencies.
The second family, {train-time calibration} \cite{lin2017focal,zhang2017mixup,yun2019cutmix,muller2019does}, modifies the training objective to explicitly penalize miscalibrated predictions.
By incorporating calibration-aware loss terms, these approaches encourage the model to produce more reliable probability estimates directly during training.
Finally, {uncertainty estimation methods} \cite{https://doi.org/10.48550/arxiv.1612.01474,pmlr-v48-gal16} focus on identifying and highlighting regions of high uncertainty, for example by quantifying epistemic or aleatoric uncertainty.
Rather than directly calibrating probabilities, these methods aim to inform practitioners about where the model’s predictions should be treated with caution, which is particularly important in high-stakes or safety-critical applications like autonomous driving.
\par
In this work, we focus exclusively on assessing the calibration capabilities of {post-hoc} methods applied to the recursive Bayesian filter introduced above.
While each family of approaches has its own advantages, {post-hoc} calibration methods offer a particularly strong benefit: they can be applied directly to pre-trained models, thereby avoiding the need to re-train potentially large and expensive architectures.
Moreover, when new data suitable for calibration become available, {post-hoc} methods provide a fast and flexible way to incorporate this additional information.
As we will show in our experiments, both the offline fitting time and the online inference time of the calibration step are low, making these methods practical even in resource-constrained or real-time settings.
This efficiency, combined with their model-agnostic nature, makes {post-hoc} calibration especially appealing for deployment scenarios where models are updated or extended frequently.
\par
In the following, let us consider a binary classification task with labels (states in our setting) $s\in\{0,1\}$ and denote by $\hat p$ the uncalibrated predicted probability for label 1.
Formally, a model is said to be perfectly calibrated if
\begin{align*}
    \mathbb{P}[s=1\mid\hat{p}]=\hat{p}.
\end{align*}
This definition formalizes what was introduced above: the true frequency of the positive class should be aligned with the model's predicted probability.
Accordingly, calibration methods aim at minimizing the distance~$|\mathbb{P}[s=1\mid\hat{p}]-\hat{p}|$.
In what follows, we first introduce three different calibration methods, and then explain how to extend these approaches to multi-class classification.

\subsubsection{Isotonic regression}
Isotonic regression~\cite{zadrozny2002transforming} is a non-parametric, monotonic calibration method that learns a piecewise constant, non-decreasing function~$f\colon [0,1] \to [0,1]$ such that the transformed scores
\begin{align*}
    p_{\text{iso}} = f(\hat{p})
\end{align*}
better match the empirical frequencies observed on a calibration set.
Formally, given~$N_\mathrm{cal}\in\mathbb{N}$ calibration pairs~$\{(\hat{p}_i, s_i)\}_{i=1}^{N_\mathrm{cal}}$, isotonic regression solves
\begin{align*}
    \min_{f \ \text{non-decreasing}}\ \sum\limits_{i=1}^{N_\mathrm{cal}} \bigl( f(\hat{p}_i) - s_i \bigr)^2,
\end{align*}
typically using the pool-adjacent-violators~(PAV) algorithm, see for instance~\cite{zadrozny2002transforming}.
The resulting function preserves the ranking of the original scores while correcting their calibration.

\subsubsection{Inductive Venn--ABERS predictor}\label{sec:venn-ABERS}
The inductive Venn--ABERS predictor~\cite{vovk2012venn} is based on the Venn prediction framework and uses isotonic regression to construct \emph{multiprobabilistic} predictions with guaranteed calibration properties under~i.i.d.\ assumptions.
For a binary task, it produces a probability interval~$\bigl[p^0_{\mathrm{VA}}, p^1_{\mathrm{VA}}\bigr]$ for label~$1$ rather than a single point estimate.
Let~$f$ be an isotonic regression model fitted on a calibration set. For a test score~$\hat{p}$, Venn--ABERS considers two hypothetical labelings:
\begin{align*}
    p^0_{\mathrm{VA}} = f\bigl(\hat{p} \mid s = 0\bigr)
    \qquad\text{and}\qquad
    p^1_{\mathrm{VA}} = f\bigl(\hat{p} \mid s = 1\bigr),
\end{align*}
and outputs the interval~$[p^0_{\mathrm{VA}}, p^1_{\mathrm{VA}}]$, which can be turned into a single probability, setting~$p_{\mathrm{VA}}=\frac{p^1_{\mathrm{VA}}}{1-p^0_{\mathrm{VA}}+p^1_{\mathrm{VA}}}$, or kept as an uncertainty-aware output.
This construction gives a form of validity: the true label frequencies are guaranteed to be compatible with the produced intervals.
Notice that, in our setting, the~i.i.d.\ assumption does not hold due to the temporal dependencies between frames.
Consequently, the theoretical validity guarantees of Venn--ABERS predictors are not strictly applicable in our scenario.
However, the construction of an interval of probabilities can still be highly informative: instead of returning a single calibrated score, the predictor provides a range~$[p^0_{\mathrm{VA}},p^1_{\mathrm{VA}}]$, which reflects the uncertainty in the underlying prediction.
In particular, a wide interval indicates that the model (and the calibration procedure) are uncertain about the true label, whereas a narrow interval suggests higher confidence.
Thus, even though the formal Venn--ABERS guarantees do not strictly hold in our application, the interval-based representation remains a useful tool for quantifying and communicating predictive uncertainty.

\subsubsection{Beta calibration}
Beta calibration~\cite{kull2017beta} is a parametric method that models the calibrated probability as a transformation of~$\hat{p}$ using a function inspired by the Beta distribution.
The formulation is
\begin{align*}
    p_{\beta}
    = \sigma\bigl(
        a \,\log \hat{p}
        + b \,\log (1 - \hat{p})
        + c
    \bigr),
\end{align*}
where~$\sigma(x) = \frac{1}{1 + e^{-x}}$ is the logistic (sigmoid) function, and~$a, b, c \in \mathbb{R}$ are parameters learned on a calibration set by minimizing a suitable loss (e.g., negative log-likelihood).
This parameterization can flexibly correct common miscalibration patterns (such as under- or over-confidence) while remaining relatively simple and efficient to train.

\subsubsection{Extension to multi-class classification}
The standard calibration methods are inherently binary.
In our setting, the labels are~$s \in \{\mathrm{K},\text{L},\mathrm{R}\}$. We are therefore dealing with a multi-class classification problem, which can be resolved by decomposing it into several binary problems.
A common strategy is the one-vs-rest~(OvR) approach~\cite{zadrozny2002transforming}. 
In this method, a single multi-class classifier is trained on all of the three classes, producing a probability vector~$(\hat p_\mathrm{K},\hat p_\mathrm{L},\hat p_\mathrm{R})$.
For each class~$s$, a binary calibration problem is then constructed by treating~$\hat p_s$ as the score for class~$s$ and pooling the remaining classes into a single ``rest'' score~$1-\hat p_s = \sum_{s' \neq s} \hat p_{s'}$.
Then, one of the proposed calibration methods, e.g.~beta calibration, is applied to this binary score~$\hat p_s$ versus the rest, yielding a calibrated one-vs-rest probability~$\tilde{p}_{\beta,s}(\hat p_s)$,
where the calibration is trained on the binary problem obtained by summing the probabilities of all classes other than~$s$ into the ``rest'' class.
The value~$\tilde{p}_{\beta,s}$ represents the calibrated probability that a sample belongs to class~$s$, given that it belongs to either class~$s$ or any of the remaining~$2$ classes.
After obtaining all three one-vs-rest probabilities, they must be converted into a valid multi-class probability distribution.
Since these scores are calibrated independently for each class, they need not sum to one; they are therefore normalized to produce the final multi-class probabilities
\begin{align*}
    {p}_{\beta,s} = \frac{\tilde{p}_{\beta,s}}{\sum\limits_{c \in \{\mathrm{K,L,R}\}}\tilde{p}_{\beta,c}}.
\end{align*}
The final predicted class~$\hat{s}$ is then selected as the class with the highest calibrated multi-class probability, that is
\begin{align*}
    \hat{s} = \arg\max_{c \in \{\mathrm{K,L,R}\}} {p}_{\beta,c}.
\end{align*}
This two-step procedure allows each calibration method to be extended to multi-class problems, first by decomposing the problem into three binary one-vs-rest splits, and then by combining the calibrated one-vs-rest probabilities into a coherent multi-class distribution~\cite{zadrozny2002transforming}.
The technique can be applied to any of the other calibration methods.
\par
A second, yet simpler, approach we can use is to binarize the problem.
In this case, the underlying model remains a three-label classifier, producing a probability vector~$(p_{\mathrm{K}}, p_{\mathrm{L}}, p_{\mathrm{R}})$.
Rather than calibrating each of the three classes individually, we merge the two lane-changing classes into a single \textit{lane changing} class, so that only one calibration problem is solved, between {lane keeping} and {lane changing}.
The score used for this binary problem is obtained by summing the probabilities of the two lane-changing classes as~$p_{\mathrm{LR}} = p_{\mathrm{L}} + p_{\mathrm{R}}$, so that~$p_{\mathrm{K}}$ and~$p_{\mathrm{LR}}$ form a valid binary distribution, i.e.~$p_{\mathrm{K}} + p_{\mathrm{LR}} = 1$.
{Post-hoc} calibration is then applied directly to this binary score, yielding a single calibrated probability.
Unlike the one-vs-rest strategy, this approach requires only a single calibration to be performed, since the three-class problem is reduced to one binary decision.
However, this comes at the cost of not providing a calibrated probability estimate for each individual class; in particular, no distinction is made between the calibrated confidence of a left versus a right lane change, as both are absorbed into the single \textit{lane changing} probability.
\par
During the experiments, we evaluate both approaches. Quantities obtained using the~OvR approach are denoted with the subscript~$\mathrm{cw}$ (class-wise), while quantities obtained using the binarized approach are denoted with the subscript~$\mathrm{lc}$ (lane changing).

\section{Experimental evaluation using the highD dataset}\label{sec:experiments}
The performance of the proposed recursive Bayesian filtering approach is evaluated on the~highD dataset.
We further compare the method to several baseline approaches.
In this section, we first describe the experimental setup and then discuss our numerical observations in detail.
\par
The implementation relies on the~\texttt{Python} programming language and makes use of the~\texttt{scikit-learn} package~\cite{scikit-learn}.
The sourcecode for reproducing our numerical results is provided in~\cite{sourcecode}.
All experiments were performed on a laptop with~Ubuntu~24.04.4 shipping an~AMD Ryzen~7 PRO 8840HS~CPU.
\par
We discuss and define the statistical metrics considered in the experiments in the appendix, see~\Cref{sec:statistical-metrics}.

\subsection{Baseline methods}\label{sec:baseline-methods}
We briefly describe three baseline methods used for comparison in the following sections.
\newline
\newline
\noindent\textbf{Lateral-velocity threshold:}
\newline
The first reference method we consider is a rule-based approach using only the lateral velocity~$v_y$ of the target vehicle.
A lane change is predicted if~$|v_y|\geq\tau$ for a threshold~$\tau$ chosen by grid search over quantiles of~$|v_y|$ in order to maximize the balanced accuracy on the training set.
This method solely takes the kinematics of the target vehicle into account and is therefore unaffected by neighbor dropout.
\newline
\newline
\noindent\textbf{Per-frame GBM:}
\newline
We consider the same~GBM as used for the filter emission and restrict it to a standalone classifier.
Each frame is labeled independently without any temporal dependencies based on the current observations.
This approach corresponds to a window length of~$w=0$ in the recursive Bayesian filter and thus allows to isolate the influence of the filtering.
\newline
\newline
\noindent\textbf{Stacked GBM:}
\newline
To also allow for incorporating potential temporal relations, we extend the per-frame~GBM to a stacked version that takes the last~$s_w$ frames of the observation as input.
Therefore, the stacked~GBM represents a learned temporal integrator, however, without the Bayesian filter.
We consider as window lengths~$s_w\in\{0.4\text{s},1.6\text{s}\}$, which correspond to~$10$ and~$40$ frames, respectively.
In the case of dropout, the~LOCF approach is used for filling the missing observation values.
The stacked~GBM provides a comparison between recursive filtering and simple stacking of the observed frames.
\newline
\newline
\noindent\textbf{TN2 from~\cite{Cristofaro2026}:}
\newline
As the final baseline we compare our results to the best-performing deep learning model in~\cite{Cristofaro2026}. The authors evaluated the Transformer-based model denoted as~TN2 using the~highD dataset, considering an observation window from~1 to~3\,s and a maximum prediction time between~3 and~6\,s. The~TN2 configuration uses multi-head attention to model temporal dependencies in the observed trajectories. The model achieved its highest accuracy of~96.73\% with class-wise~$F_1$ scores of~96,80\% at a~3\,s observation window and maximum prediction time. Based on the overall comparison in the study, TN2 was reported as the best-performing model. We apply the same evaluation strategy including the same metrics.
However, we note that our evaluation set most likely differs from the one used in~\cite{Cristofaro2026} due to missing access to their implementation.

\subsection{Numerical results}
In the following, the experimental results are presented.
The section is divided into several parts, devoted to the analysis of different components of the recursive Bayesian filtering approach.
If not stated differently, we use the compact~16-features observation set, $h=4$\,s and~$w=0.4$\,s.

\subsubsection{Feature selection on uncorrupted data}\label{sec:feature-selection}
We start our investigation by examining the influence of the different groups of features on the prediction capabilities of the individual models.
The guiding question is if taking into account surrounding features, besides the kinematic features, improves the accuracy of the recursive Bayesian filtering.
\par
\begin{figure}[htbp]
  \centering
  \begin{tikzpicture}
  \begin{groupplot}[group style={group size=2 by 1, horizontal sep=1.15cm}, paperaxis, ylabel={balanced accuracy (\%)}, ymin=69.0, ymax=101.4]
    \nextgroupplot[title={GMM}, xlabel={prediction horizon $h$ (s)}, xtick={1,2,3,4,5,6}, legend style={at={(1.075,-0.22)}, anchor=north, yshift=-16pt, legend columns=-1, /tikz/every even column/.append style={column sep=6pt}}]
      \addplot[kinematics] table[x=H, y=kinematics]{\datpath scope_horizon_gmm.dat};
      \addplot[compact] table[x=H, y=compact]{\datpath scope_horizon_gmm.dat};
      \addplot[fullCont] table[x=H, y=fullCont]{\datpath scope_horizon_gmm.dat};
      \addplot[fullPresent] table[x=H, y=fullPresent]{\datpath scope_horizon_gmm.dat};
      \legend{kinematics (4), compact (16), full-cont.~(20), full+flags (28)}
    \nextgroupplot[title={GBM}, xlabel={prediction horizon $h$ (s)}, xtick={1,2,3,4,5,6}, ylabel={}]
      \addplot[kinematics] table[x=H, y=kinematics]{\datpath scope_horizon_gbm.dat};
      \addplot[compact] table[x=H, y=compact]{\datpath scope_horizon_gbm.dat};
      \addplot[fullCont] table[x=H, y=fullCont]{\datpath scope_horizon_gbm.dat};
      \addplot[fullPresent] table[x=H, y=fullPresent]{\datpath scope_horizon_gbm.dat};
  \end{groupplot}
  \end{tikzpicture}
  \caption{Balanced accuracy across different prediction horizons for~GMM- and~GBM-based methods using different feature scopes.}
  \label{fig:scope}
\end{figure}
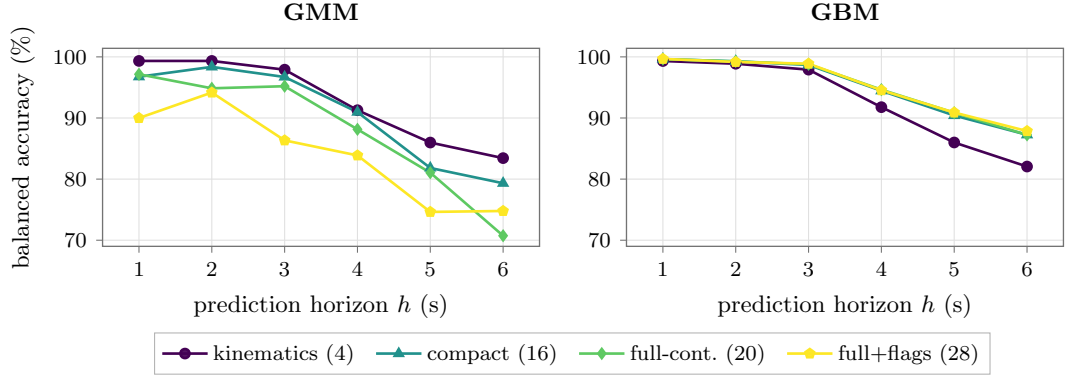

In~\Cref{fig:scope}, we show the balanced accuracy of the~GMM- and~GBM-based emission models, respectively, depending on the prediction horizon~$h$.
For both methods, the accuracy decreases when increasing the prediction horizon.
As expected, having to predict a lane change in advance is more difficult for the considered models.
\par
We observe that for the~GMM, reducing the set of features to the four kinematic observations improves the accuracy independent of the considered horizon.
The main reason for this behavior is the difficulty of dealing with high-dimensional observations, as mentioned in~\Cref{rem:GMMs-high-dimensions}.
This behavior will become apparent also in our later experiments, where dropout of surrounding features is actually beneficial for the~GMM.
However, we will also observe in~\Cref{sec:robustness-against-dropout} that the marginalization of the missing observations improves the accuracy compared to a model that only sees the kinematic features in the first place.
\par
For the~GBM-based approach, adding surrounding features generally improves the performance, in particular for longer prediction horizons.
However, adding features beyond the compact feature set (containing~$16$ observations in total) does not pay off and might even reduce the accuracy.
We therefore restrict our attention to the compact feature set in the subsequent sections.

\subsubsection{Choice of training data including dropout}
In this section we evaluate to what extend augmenting the training data by observation dropout improves the performance under~i.i.d.\ and bursty dropout during inference.
\par
\begin{figure}[htbp]
  \centering
  \begin{tikzpicture}
  \begin{groupplot}[group style={group size=2 by 1, horizontal sep=1.15cm}, paperaxis, ylabel={balanced accuracy (\%)}, ymin=84.6, ymax=95.1]
    \nextgroupplot[title={i.i.d.\ dropout}, xlabel={neighbour dropout $\rho$}, legend columns=5, legend cell align=left, legend style={at={(1.08,-0.22)}, anchor=north, yshift=-16pt, /tikz/every even column/.append style={column sep=5pt}}]
      \addplot[gmmMarg, dashed, forget plot] table[x=rho, y=gmmMarg]{\datpath train_dropout_iid.dat};
      \addplot[gbmSentClean, forget plot] table[x=rho, y=gbmSentClean]{\datpath train_dropout_iid.dat};
      \addplot[gbmMeanClean, forget plot] table[x=rho, y=gbmMeanClean]{\datpath train_dropout_iid.dat};
      \addplot[gbmLOCFClean, forget plot] table[x=rho, y=gbmLOCFClean]{\datpath train_dropout_iid.dat};
      \addplot[gbmSentAug, forget plot] table[x=rho, y=gbmSentAug]{\datpath train_dropout_iid.dat};
      \addplot[gbmMeanAug, forget plot] table[x=rho, y=gbmMeanAug]{\datpath train_dropout_iid.dat};
      \addplot[gbmLOCFAug, forget plot] table[x=rho, y=gbmLOCFAug]{\datpath train_dropout_iid.dat};
      \addlegendimage{empty legend}\addlegendentry{clean}
      \addlegendimage{gmmMarg, dashed}\addlegendentry{GMM-marg}
      \addlegendimage{gbmSentClean}\addlegendentry{GBM-sentinel}
      \addlegendimage{gbmMeanClean}\addlegendentry{GBM-mean}
      \addlegendimage{gbmLOCFClean}\addlegendentry{GBM-LOCF}
      \addlegendimage{empty legend}\addlegendentry{augmented}
      \addlegendimage{empty legend}\addlegendentry{}
      \addlegendimage{gbmSentAug}\addlegendentry{GBM-sentinel}
      \addlegendimage{gbmMeanAug}\addlegendentry{GBM-mean}
      \addlegendimage{gbmLOCFAug}\addlegendentry{GBM-LOCF}
    \nextgroupplot[title={bursty dropout}, xlabel={neighbour dropout $\rho$}, ylabel={}]
      \addplot[gmmMarg, dashed, forget plot] table[x=rho, y=gmmMarg]{\datpath train_dropout_bursty.dat};
      \addplot[gbmSentClean, forget plot] table[x=rho, y=gbmSentClean]{\datpath train_dropout_bursty.dat};
      \addplot[gbmMeanClean, forget plot] table[x=rho, y=gbmMeanClean]{\datpath train_dropout_bursty.dat};
      \addplot[gbmLOCFClean, forget plot] table[x=rho, y=gbmLOCFClean]{\datpath train_dropout_bursty.dat};
      \addplot[gbmSentAug, forget plot] table[x=rho, y=gbmSentAug]{\datpath train_dropout_bursty.dat};
      \addplot[gbmMeanAug, forget plot] table[x=rho, y=gbmMeanAug]{\datpath train_dropout_bursty.dat};
      \addplot[gbmLOCFAug, forget plot] table[x=rho, y=gbmLOCFAug]{\datpath train_dropout_bursty.dat};
  \end{groupplot}
  \end{tikzpicture}
  \caption{Comparison of different methods trained on clean and augmented training data.}
  \label{fig:traindropout}
\end{figure}
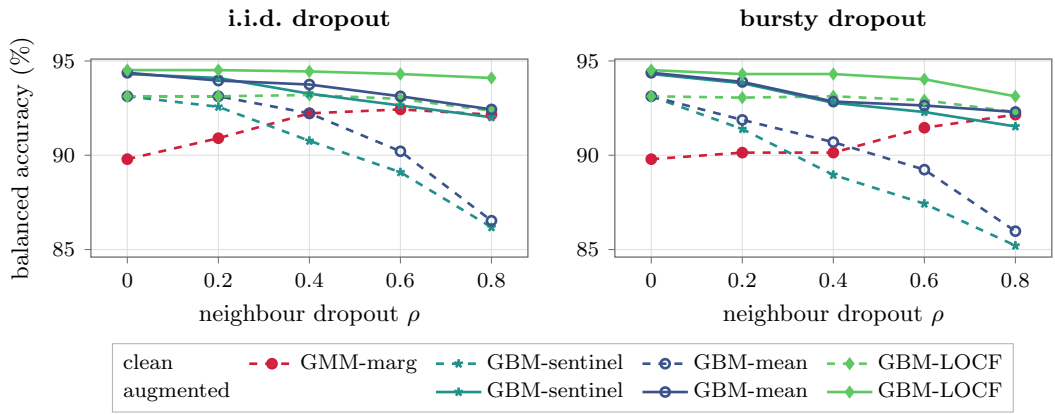

\Cref{fig:traindropout} presents the balanced accuracy for the~GMM approach with marginalization (referred to as~``GMM-marg'') and different variants of the~GBM-based emission model using sentinel, mean and~LOCF dropout imputation.
The figure distinguishes models trained on clean data and those that saw augmented data during training, based on the augmentation process described at the end of~\Cref{sec:model-observation-dropout}.
\par
We observe that in particular the~GBM-sentinel and~GBM-mean models benefit significantly from seeing observation dropout already during training, in particular for larger dropout rates.
The accuracy of the~GBM-LOCF can also be improved by augmenting the training data, however, also when employing clean data during training the model remains accurate independent of the dropout rate.
The behavior is similar for both~i.i.d.\ and bursty dropout.
For the~GMM-marg model, it turns out that higher dropout rates lead to higher accuracy, finally even matching the~GBM-sentinel and~GBM-mean models for a dropout rate of~$\rho=0.8$.
As discussed in the previous section, this behavior can be explained by the difficulties of~GMMs in high dimensions and the reduction of the observation dimension via dropout.
The results in~\Cref{fig:traindropout} also show that using~LOCF for imputation leads to the best results in terms of balanced accuracy, independent of the dropout rate and the dropout mechanism.
\par
Based on the findings from this section, all~GBM-based models below will be trained on augmented training data.
Only the~GMM-marg model will see the entire, clean data for training.

\subsubsection{Robustness against dropout}\label{sec:robustness-against-dropout}
We now first focus on analyzing the methods' robustness against dropout by considering different dropout rates, prediction horizons, and the statistical metrics introduced in~\Cref{sec:statistical-metrics}..
\par
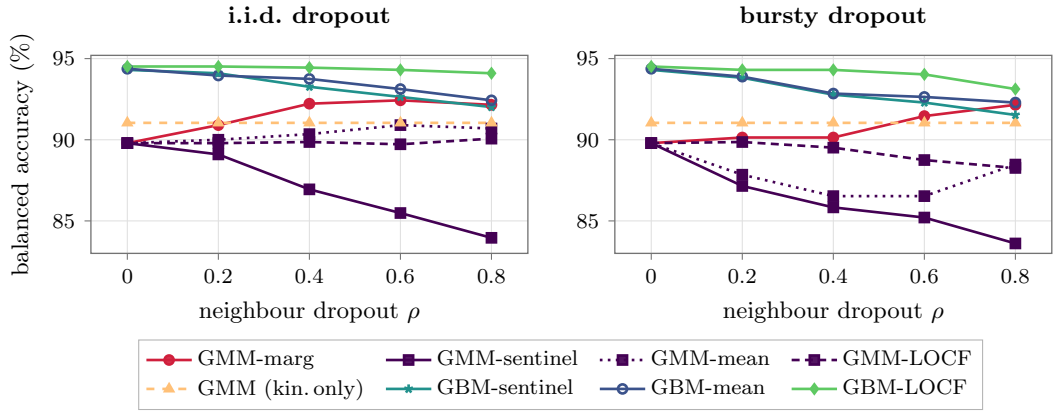
\begin{figure}[htbp]
  \centering
  \begin{tikzpicture}
  \begin{groupplot}[group style={group size=2 by 1, horizontal sep=1.15cm}, paperaxis, ylabel={balanced accuracy (\%)}, ymin=83.0, ymax=95.2]
    \nextgroupplot[title={i.i.d.\ dropout}, xlabel={neighbour dropout $\rho$}, legend style={at={(1.075,-0.22)}, anchor=north, yshift=-16pt, legend columns=4, /tikz/every even column/.append style={column sep=6pt}}]
      \addplot[gmmMarg] table[x=rho, y=gmmMarg]{\datpath robustness_dropout_iid.dat};
      \addplot[gmmSent] table[x=rho, y=gmmSent]{\datpath robustness_dropout_iid.dat};
      \addplot[gmmMean] table[x=rho, y=gmmMean]{\datpath robustness_dropout_iid.dat};
      \addplot[gmmLOCF] table[x=rho, y=gmmLOCF]{\datpath robustness_dropout_iid.dat};
      \addplot[gmmKin] table[x=rho, y=gmmKin]{\datpath robustness_dropout_iid.dat};
      \addplot[gbmSent] table[x=rho, y=gbmSent]{\datpath robustness_dropout_iid.dat};
      \addplot[gbmMean] table[x=rho, y=gbmMean]{\datpath robustness_dropout_iid.dat};
      \addplot[gbmLOCF] table[x=rho, y=gbmLOCF]{\datpath robustness_dropout_iid.dat};
      \legend{GMM-marg, GMM-sentinel, GMM-mean, GMM-LOCF, GMM (kin.\,only), GBM-sentinel, GBM-mean, GBM-LOCF}
    \nextgroupplot[title={bursty dropout}, xlabel={neighbour dropout $\rho$}, ylabel={}]
      \addplot[gmmMarg] table[x=rho, y=gmmMarg]{\datpath robustness_dropout_bursty.dat};
      \addplot[gmmSent] table[x=rho, y=gmmSent]{\datpath robustness_dropout_bursty.dat};
      \addplot[gmmMean] table[x=rho, y=gmmMean]{\datpath robustness_dropout_bursty.dat};
      \addplot[gmmLOCF] table[x=rho, y=gmmLOCF]{\datpath robustness_dropout_bursty.dat};
      \addplot[gmmKin] table[x=rho, y=gmmKin]{\datpath robustness_dropout_bursty.dat};
      \addplot[gbmSent] table[x=rho, y=gbmSent]{\datpath robustness_dropout_bursty.dat};
      \addplot[gbmMean] table[x=rho, y=gbmMean]{\datpath robustness_dropout_bursty.dat};
      \addplot[gbmLOCF] table[x=rho, y=gbmLOCF]{\datpath robustness_dropout_bursty.dat};
  \end{groupplot}
  \end{tikzpicture}
  \caption{Robustness to partial neighbor dropout at fixed prediction horizon~$h=4$\,s.}
  \label{fig:robustness}
\end{figure}
\begin{figure}[htbp]
  \centering
  \begin{tikzpicture}
  \begin{groupplot}[group style={group size=2 by 1, horizontal sep=1.15cm}, paperaxis, ylabel={balanced accuracy (\%)}, ymin=68.2, ymax=101.7]
    \nextgroupplot[title={i.i.d.\ dropout}, xlabel={prediction horizon $h$ (s)}, xtick={1,2,3,4,5,6}, legend style={at={(1.075,-0.22)}, anchor=north, yshift=-16pt, legend columns=4, /tikz/every even column/.append style={column sep=6pt}}]
      \addplot[gmmMarg] table[x=H, y=gmmMarg]{\datpath robustness_horizon_iid.dat};
      \addplot[gmmSent] table[x=H, y=gmmSent]{\datpath robustness_horizon_iid.dat};
      \addplot[gmmMean] table[x=H, y=gmmMean]{\datpath robustness_horizon_iid.dat};
      \addplot[gmmLOCF] table[x=H, y=gmmLOCF]{\datpath robustness_horizon_iid.dat};
      \addplot[gmmKin] table[x=H, y=gmmKin]{\datpath robustness_horizon_iid.dat};
      \addplot[gbmSent] table[x=H, y=gbmSent]{\datpath robustness_horizon_iid.dat};
      \addplot[gbmMean] table[x=H, y=gbmMean]{\datpath robustness_horizon_iid.dat};
      \addplot[gbmLOCF] table[x=H, y=gbmLOCF]{\datpath robustness_horizon_iid.dat};
      \legend{GMM-marg, GMM-sentinel, GMM-mean, GMM-LOCF, GMM (kin.\,only), GBM-sentinel, GBM-mean, GBM-LOCF}
    \nextgroupplot[title={bursty dropout}, xlabel={prediction horizon $h$ (s)}, xtick={1,2,3,4,5,6}, ylabel={}]
      \addplot[gmmMarg] table[x=H, y=gmmMarg]{\datpath robustness_horizon_bursty.dat};
      \addplot[gmmSent] table[x=H, y=gmmSent]{\datpath robustness_horizon_bursty.dat};
      \addplot[gmmMean] table[x=H, y=gmmMean]{\datpath robustness_horizon_bursty.dat};
      \addplot[gmmLOCF] table[x=H, y=gmmLOCF]{\datpath robustness_horizon_bursty.dat};
      \addplot[gmmKin] table[x=H, y=gmmKin]{\datpath robustness_horizon_bursty.dat};
      \addplot[gbmSent] table[x=H, y=gbmSent]{\datpath robustness_horizon_bursty.dat};
      \addplot[gbmMean] table[x=H, y=gbmMean]{\datpath robustness_horizon_bursty.dat};
      \addplot[gbmLOCF] table[x=H, y=gbmLOCF]{\datpath robustness_horizon_bursty.dat};
  \end{groupplot}
  \end{tikzpicture}
  \caption{Balanced accuracy versus prediction horizon at fixed dropout rate~$\rho=0.6$.}
  \label{fig:robhorizon}
\end{figure}
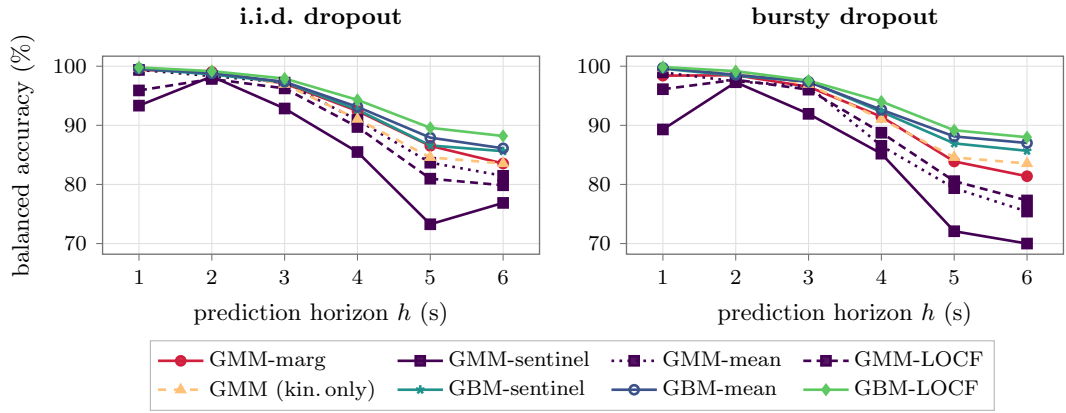

In~\Cref{fig:robustness,fig:robhorizon}, we also report the accuracy for~GMM-based emission models using sentinel, mean and~LOCF imputation.
We only show the results obtained when training on clean data.
Augmenting the data improves the results slightly while still remaining below the~GMM-marg model in accuracy.
The reason for including those imputation-based models as well is to show that using marginalization (which is the natural way to handle missing observations for a generative model such as the~GMM) leads to a more accurate method compared to imputing.
\par
\Cref{fig:robustness} shows the balanced accuracy at a fixed prediction horizon of~$h=4$\,s under different neighbor dropout rates~$\rho$.
As mentioned above, the~GMM-marg model has a higher accuracy than all of the other~GMM-based models (except for the~GMM using only kinematic features for small dropout rates).
The~GMM-sentinel model degrades significantly at higher dropout rates whereas all other models either improve or roughly retain their accuracy.
The~GMM based solely on kinematic observations of the target vehicle is unaffected by dropout such that its performance remains constant over the different dropout rates and types.
As observed previously, the~GBM-based methods achieve higher accuracy compared to the~GMM-based variants.
\par
We now fix a dropout rate of~$\rho=0.6$ and investigate the balanced accuracy of the discussed approaches with respect to the prediction horizon~$h\in\{1,\ldots,6\}$.
First of all, we observe that the general trends are independent of the dropout mechanism, i.e.~if~i.i.d.\ or bursty dropout is considered at test time.
For longer prediction horizons, the accuracy decreases by more than~$10\%$ for all methods.
The~GMM-based methods generally perform worse than their corresponding~GBM emission models using the same imputation strategy.
Only the~GMM-marg approach remains roughly on par with the~GBM variants up to a prediction of~$h=5$\,s for~i.i.d.\ dropout and~$h=4$\,s for bursty dropout.
We also see that bursty dropout leads to a significantly reduced accuracy for the~GMM-based methods, whereas the~GBM-based approaches are similarly affected by~i.i.d.\ and bursty dropout.
As before, the~GBM-LOCF technique yields the best accuracy among all considered methods, independent of the prediction horizon and the dropout mechanism.
\par
In~\Cref{tab:robustness} we present the results of the~GMM-marg, GMM with kinematic features only and the different~GBM variants in terms of several statistical metrics as introduced in~\Cref{sec:statistical-metrics}.
We further investigate different dropout rates ranging from clean observations over partial neighbor dropout at rate~$\rho=0.4$ to severe dropout for~$\rho=0.8$.
\par
In terms of accuracy, we observe that the~GBM-based methods in general lead to higher accuracies compared to the~GMM-based approaches.
In particular~GBM-LOCF is the most accurate independent of the dropout rate.
For a strong dropout with rate~$\rho=0.8$, the~GMM-marg matches the balanced accuracy of the other methods.
Also for the recall, the~GBM-LOCF yields the best results.
Overall, the recall of the keep-lane class is the highest for all methods.
For the precision, however, lane change left and right show high precision values.
The~macro-$F_1$ metric is almost above~90\% for all considered methods at all dropout rates.
\begin{table}[htbp]
  \centering
  \begin{tabular}{lcccccccc}
    \toprule
    Method & $\mathrm{bAcc}$ & $\mathrm{rec}_\mathrm{K}$ & $\mathrm{rec}_\mathrm{L}$ & $\mathrm{rec}_\mathrm{R}$ & $\mathrm{pr}_\mathrm{K}$ & $\mathrm{pr}_\mathrm{L}$ & $\mathrm{pr}_\mathrm{R}$ & $F_1^{\mathrm{macro}}$ \\
    \midrule
    \multicolumn{9}{l}{\emph{dropout $\rho=0$}} \\
    GMM-marg & 89.8 & 92.9 & 88.8 & 87.7 & 81.5 & 95.7 & 94.0 & 89.9 \\
    GMM (kin.\,only) & 91.0 & 96.9 & 89.6 & 86.7 & 80.3 & 97.7 & 98.8 & 91.2 \\
    GBM-sentinel & 94.3 & 97.5 & 93.8 & 91.7 & 87.0 & 99.3 & 98.0 & 94.4 \\
    GBM-mean & 94.4 & 98.1 & 93.8 & 91.2 & 86.7 & 99.8 & 98.2 & 94.5 \\
    GBM-LOCF & 94.5 & 98.3 & 94.4 & 90.8 & 87.6 & 98.7 & 98.6 & 94.6 \\
    \multicolumn{9}{l}{\emph{dropout $\rho=0.4$}} \\
    GMM-marg & 92.2 & 97.5 & 91.2 & 87.9 & 82.5 & 98.6 & 98.4 & 92.4 \\
    GMM (kin.\,only) & 91.0 & 96.9 & 89.6 & 86.7 & 80.3 & 97.7 & 98.8 & 91.2 \\
    GBM-sentinel & 93.3 & 96.7 & 91.2 & 91.9 & 85.1 & 98.9 & 97.6 & 93.4 \\
    GBM-mean & 93.8 & 97.5 & 92.3 & 91.5 & 85.7 & 99.3 & 98.0 & 93.8 \\
    GBM-LOCF & 94.4 & 98.1 & 94.4 & 90.8 & 87.5 & 98.7 & 98.4 & 94.5 \\
    \multicolumn{9}{l}{\emph{dropout $\rho=0.8$}} \\
    GMM-marg & 92.2 & 97.7 & 90.6 & 88.1 & 82.1 & 98.6 & 98.8 & 92.3 \\
    GMM (kin.\,only) & 91.0 & 96.9 & 89.6 & 86.7 & 80.3 & 97.7 & 98.8 & 91.2 \\
    GBM-sentinel & 92.0 & 96.5 & 89.8 & 89.8 & 82.5 & 98.4 & 97.7 & 92.1 \\
    GBM-mean & 92.4 & 96.5 & 90.4 & 90.4 & 83.4 & 98.6 & 97.5 & 92.6 \\
    GBM-LOCF & 94.1 & 97.5 & 94.6 & 90.2 & 87.2 & 98.7 & 97.7 & 94.2 \\
    \bottomrule
  \end{tabular}
  \caption{Balanced accuracy, per-class recall, per-class precision and macro-$F_1$ under partial neighbor dropout (compact-16; balanced eval; $h=4$\,s; $\rho\in\{0,0.4,0.8\}$; 3 seeds; all in \%). GMM+imputation variants (sentinel/mean/LOCF) are dominated by marginalisation and are shown only in Fig.~\ref{fig:robustness}. Belief calibration is studied separately in Table~\ref{tab:calibration}.}
  \label{tab:robustness}
\end{table}

\subsubsection{Temporal consistency}\label{sec:temporal-consistency}
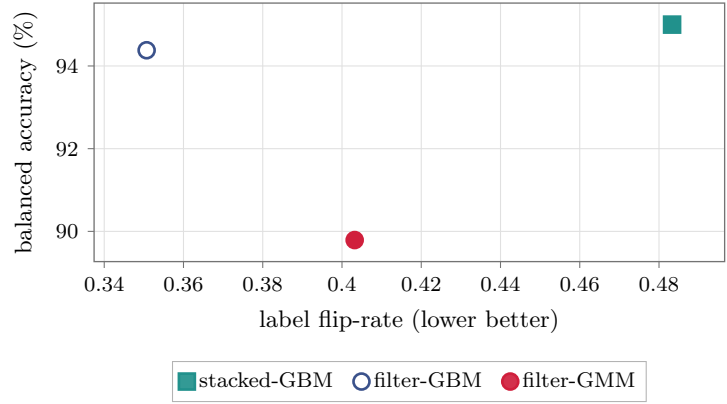
\begin{figure}[htbp]
  \centering
  \begin{tikzpicture}
  \begin{axis}[paperaxis, width=0.62\linewidth, height=5cm, xlabel={label flip-rate (lower better)}, ylabel={balanced accuracy (\%)}, legend style={at={(0.5,-0.22)}, anchor=north, yshift=-16pt, legend columns=-1, /tikz/every even column/.append style={column sep=6pt}}]
    \addplot[stackedGBM, only marks, mark size=3pt] coordinates {(0.4833,95.00)};
    \addlegendentry{stacked-GBM}
    \addplot[filterGBM, only marks, mark size=3pt] coordinates {(0.3507,94.38)};
    \addlegendentry{filter-GBM}
    \addplot[filterGMM, only marks, mark size=3pt] coordinates {(0.4032,89.79)};
    \addlegendentry{filter-GMM}
  \end{axis}
  \end{tikzpicture}
  \caption{Temporal consistency in terms of flip-rate versus accuracy on clean test data. All methods use the same compact feature set containing~16 observations and operate on a window length of~$w=0.4$\,s.}
  \label{fig:consistency}
\end{figure}

After assessing the performance of the different approaches in terms of accuracy, recall, and precision, we now turn our attention towards temporal consistency.
To this end, \Cref{fig:consistency} shows the flip-rate compared to the balanced accuracy for the stacked-GBM variant (which does not involve the recursive Bayesian filter) and the filter-based~GBM and~GMM methods (to facilitate distinction we explicitly mention ``filter'' in the legend of the plot --  the results shown there refer to the~GBM-LOCF and~GMM-marg methods from before).
In this experiment, we used clean data for testing in order to focus on temporal consistency of the predictions rather than dealing with missing observations.
We observe that both the~filter-GBM and the~filter-GMM methods lead to a lower flip-rate.
In particular the~filter-GBM reduces the flip-rate by roughly one third while obtaining an accuracy comparable to the~stacked-GBM version.
In cases where temporal consistency matter, the recursive filter thus provides more reliable predictions at a similar level of accuracy.
\par
\begin{figure}[htbp]
  \centering
  \begin{tikzpicture}
  \begin{groupplot}[group style={group size=3 by 1, horizontal sep=1.3cm}, paperaxis, width=0.29\linewidth, xlabel={filter window $w$ (s)}, xtick={0,1,2,3,4,5,6}, xticklabels={0,0.2,0.4,0.8,1.6,3.2,$\infty$}, x tick label style={font=\scriptsize}]
    \nextgroupplot[title={accuracy}, ylabel={balanced accuracy (\%)}, legend style={at={(1.65,-0.22)}, anchor=north, yshift=-16pt, legend columns=-1, /tikz/every even column/.append style={column sep=6pt}}]
      \addplot[filterGMM] table[x=idx, y=filterGMM]{\datpath window_acc.dat};
      \addplot[filterGBM] table[x=idx, y=filterGBM]{\datpath window_acc.dat};
      \legend{filter-GMM, filter-GBM}
    \nextgroupplot[title={temporal consistency}, ylabel={flip-rate (\%)}]
      \addplot[filterGMM] table[x=idx, y=filterGMM]{\datpath window_flip.dat};
      \addplot[filterGBM] table[x=idx, y=filterGBM]{\datpath window_flip.dat};
    \nextgroupplot[title={anticipation}, ylabel={recall@1\,s (\%)}]
      \addplot[filterGMM] table[x=idx, y=filterGMM]{\datpath window_recall.dat};
      \addplot[filterGBM] table[x=idx, y=filterGBM]{\datpath window_recall.dat};
  \end{groupplot}
  \end{tikzpicture}
  \caption{Recursive Bayesian filter performance under different filter window lengths~$w$, evaluated on clean test data. The rightmost plot shows the sustained recall for a required lead time of~$d=1$\,s.}
  \label{fig:window}
\end{figure}
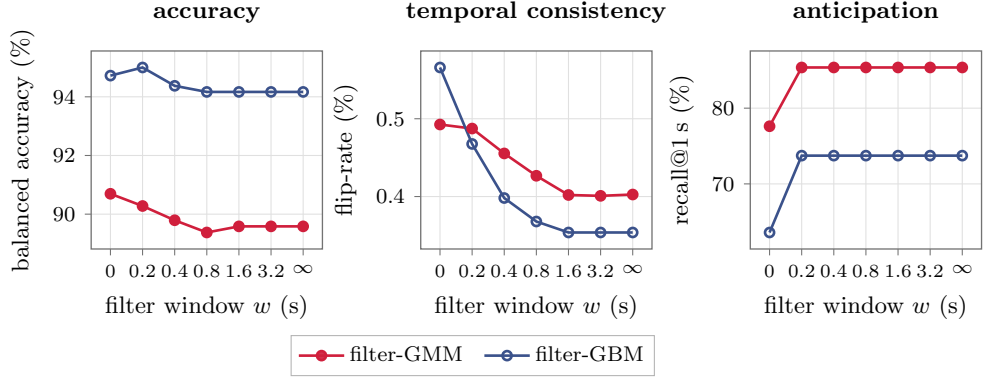

We further compare in~\Cref{fig:window} the~filter-GMM and~filter-GBM methods in terms of accuracy, temporal consistency and anticipation capabilities when changing the filter window length~$w$.
Here, filter windows of length~$w=0$\,s up to~$w=3.2$\,s, and also unbounded filters, i.e.~$w=\infty$, are considered.
The numerical experiments indicate that the accuracy decreases only slightly when increasing the filter window length.
In general, a longer filter window leads to stronger smoothing of the probabilities in time, which affects the accuracy of the predictions.
However, this effect seems to be rather limited in our case.
The flip-rate, however, is significantly reduced when using a longer filter window.
The~GBM benefits here in particular from longer filter windows while loosing less than~$1\%$ of accuracy when moving from~$w=0$ to~$w=\infty$.
In~\Cref{sec:statistical-metrics} we also introduced the recall at required lead~$d$, which measures the anticipation capabilities of the recursive Bayesian filter by requiring it to commit early to a certain prediction.
Here, we focus on sustained commitment to a prediction and observe that the recall improves considerably when introducing a positive filter window length.
Even for a relatively small value of~$w=0.2$\,s, the filter-based methods seem to anticipate lane changes about one second ahead much more often than the unfiltered versions (with~$w=0$).
\par
\begin{figure}[htbp]
  \centering
  \begin{tikzpicture}
  \begin{axis}[paperaxis, width=0.62\linewidth, height=5cm, xlabel={filter window $w$ (s)}, ylabel={balanced accuracy (\%)}, xtick={0,1,2,3,4,5,6}, xticklabels={0,0.2,0.4,0.8,1.6,3.2,$\infty$}, legend style={at={(0.5,-0.22)}, anchor=north, yshift=-16pt, legend columns=-1, /tikz/every even column/.append style={column sep=6pt}}]
    \addplot[color=viridisViolet, solid, mark=*] table[x=idx, y=clean]{\datpath window_dropout.dat};
    \addlegendentry{clean}
    \addplot[color=viridisBlue, solid, mark=square*] table[x=idx, y=neighbour]{\datpath window_dropout.dat};
    \addlegendentry{neighbor dropout ($\rho{=}0.8$)}
    \addplot[color=matchingRed, solid, mark=diamond*] table[x=idx, y=blackout]{\datpath window_dropout.dat};
    \addlegendentry{full blackout ($\rho{=}0.5$)}
  \end{axis}
  \end{tikzpicture}
  \caption{Filter window under occlusion for the~GMM-marg model under bursty observation dropout. Besides pure neighbor dropout, the plot also includes results for full blackout in which the kinematic features have been removed as well.}
  \label{fig:windowdropout}
\end{figure}
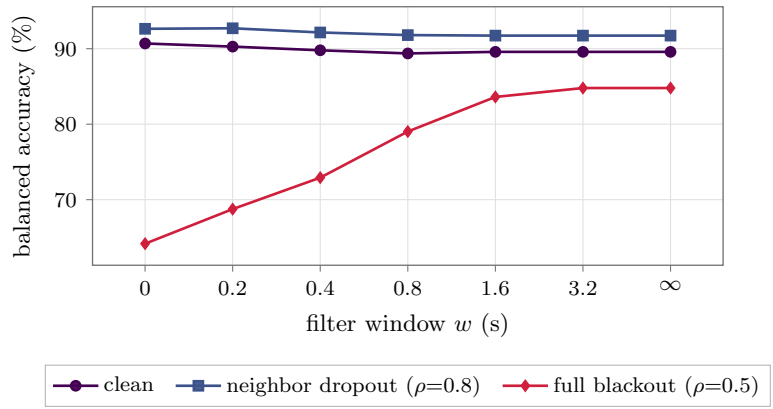

In~\Cref{fig:windowdropout} we investigate the~GMM-marg model under different dropout types with respect to the filter window length~$w$.
For clean and severe neighbor dropout at rate~$\rho=0.8$, the balanced accuracy is almost constant, independent of the filter window size.
However, if we additionally consider a full blackout, i.e.~also kinematic features are not available as observations, the accuracy improves significantly from roughly~65\% to around~85\% when using a filter window length of~$w=3.2$\,s or~$w=\infty$.
Choosing a larger window size~$w$ for the~GMM-marg approach is therefore in particular preferable when also an entire blackout of the observations is to be expected.
\par
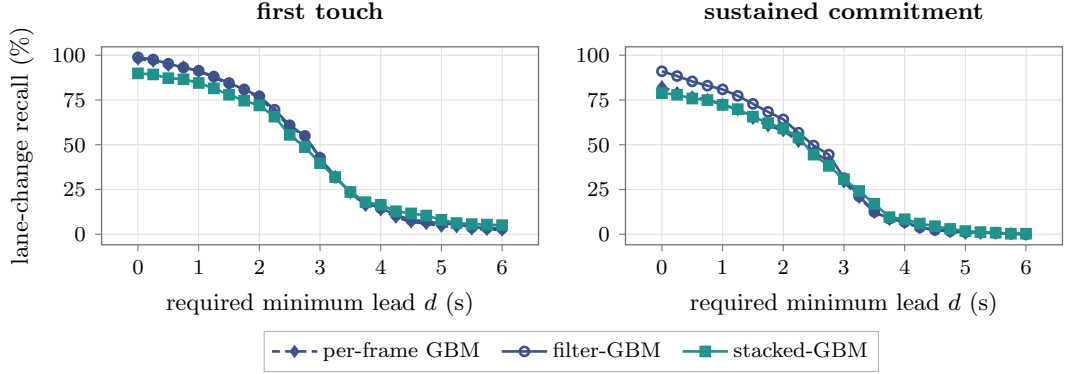
\begin{figure}[htbp]
  \centering
  \begin{tikzpicture}
  \begin{groupplot}[group style={group size=2 by 1, horizontal sep=1.15cm}, paperaxis, ylabel={lane-change recall (\%)}, ymin=-5.9, ymax=104.7]
    \nextgroupplot[title={first touch}, xlabel={required minimum lead $d$ (s)}, xtick={0,1,2,3,4,5,6}, ytick={0,25,50,75,100}, legend style={at={(1.075,-0.22)}, anchor=north, yshift=-16pt, legend columns=-1, /tikz/every even column/.append style={column sep=6pt}}]
      \addplot[perFrameGBM] table[x=d, y=perFrameGBM]{\datpath lead_firsttouch.dat};
      \addplot[filterGBM] table[x=d, y=filterGBM]{\datpath lead_firsttouch.dat};
      \addplot[stackedGBM] table[x=d, y=stackedGBM]{\datpath lead_firsttouch.dat};
      \legend{per-frame GBM, filter-GBM, stacked-GBM}
    \nextgroupplot[title={sustained commitment}, xlabel={required minimum lead $d$ (s)}, xtick={0,1,2,3,4,5,6}, ytick={0,25,50,75,100}, ylabel={}]
      \addplot[perFrameGBM] table[x=d, y=perFrameGBM]{\datpath lead_sustained.dat};
      \addplot[filterGBM] table[x=d, y=filterGBM]{\datpath lead_sustained.dat};
      \addplot[stackedGBM] table[x=d, y=stackedGBM]{\datpath lead_sustained.dat};
  \end{groupplot}
  \end{tikzpicture}
  \caption{Detection versus anticipation capabilities of the~per-frame~GBM and the~filter-GBM models: recall of true lane changes when the correct side must be flagged at least~$d$\,s before the crossing. The models were trained at~$h=6$\,s and evaluated on clean data. \emph{Left}: first-touch (the score needs to cross the threshold only once). \emph{Right}: sustained commitment (the correct side must stay above the threshold through the crossing).}
  \label{fig:anticipation}
\end{figure}

We finally analyze the lane-change recall for different required minimum lead times~$d$.
\Cref{fig:anticipation} shows the performance of per-frame~GBM, filter-GBM and stacked-GBM for first-touch and sustained commitment.
The numerical results show that if a single cross of the detection threshold is sufficient (first touch), the per-frame~GBM and filter-GBM perform slightly better for small values of~$d$, whereas the stacked-GBM shows a higher recall (although at very low level, as expected) for long required lead times.
For the sustained commitment case where the model has to predict correctly for the entire interval of length~$d$ before the actual crossing happens, the per-frame~GBM and the stacked-GBM perform similar.
The filter-GBM, however, yields a higher lane-change recall for values of~$d$ up to around~$3$\,s.
The recursive filter thus seems to facilitate early commitment to a lane-change prediction and therefore improves temporal consistency of the forecasts.

\subsubsection{Calibration}\label{sec:calibration-results}
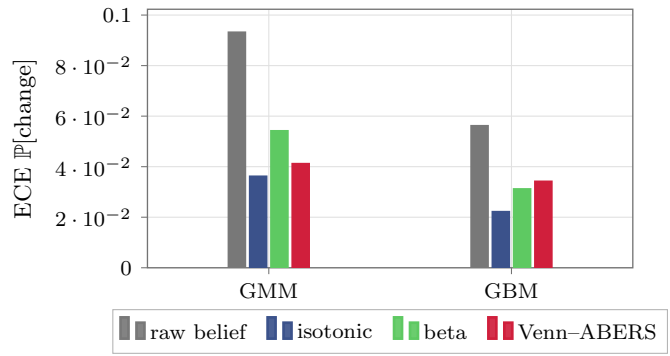
\begin{figure}[htbp]
  \centering
  \begin{tikzpicture}
  \begin{axis}[paperaxis, width=0.5\linewidth, height=5cm, ybar, bar width=6pt, ymin=0, enlarge x limits=0.5, ylabel={ECE $\mathbb{P}[\text{change}]$}, symbolic x coords={GMM,GBM}, xtick=data, legend style={at={(0.5,-0.16)}, anchor=north, legend columns=-1, /tikz/every even column/.append style={column sep=6pt}}]
    \addplot[fill=refGray, draw=refGray] coordinates {(GMM,0.093) (GBM,0.056)};
    \addlegendentry{raw belief}
    \addplot[fill=viridisBlue, draw=viridisBlue] coordinates {(GMM,0.036) (GBM,0.022)};
    \addlegendentry{isotonic}
    \addplot[fill=viridisGreen, draw=viridisGreen] coordinates {(GMM,0.054) (GBM,0.031)};
    \addlegendentry{beta}
    \addplot[fill=matchingRed, draw=matchingRed] coordinates {(GMM,0.041) (GBM,0.034)};
    \addlegendentry{Venn--ABERS}
  \end{axis}
  \end{tikzpicture}
  \caption{Calibration error of the filtered belief comparing the raw belief to the three calibrators (isotonic regression, beta calibration, and Venn--ABERS) for the binary probability~$\mathbb{P}[\text{change}]$. Here we are considering the ECE$_{\mathrm{lc}}$.}
  \label{fig:calibration}
\end{figure}

We now discuss the effect of different calibration methods.
In~\Cref{fig:calibration} we show the~ECE$_{\mathrm{lc}}$ for the~GMM and the~GBM models, and in particular we illustrate the effect of the different calibration methods.
First of all, we can notice that even the uncalibrated models reach a relatively small calibration error of less than~$10\%$, especially the~GBM, with an error lower than~$6\%$.
This is probably due to the much lower capacity of these models compared to standard deep-learning models.
Nevertheless, it is important to reduce this calibration error as much as possible, and the calibration methods fulfill that purpose.
All three methods reduce the~ECE$_{\mathrm{lc}}$, with isotonic regression being the best one for both models.
\par
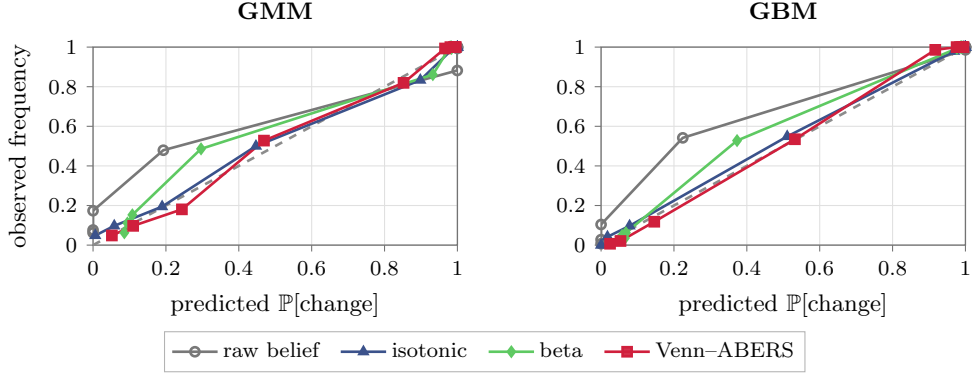
\begin{figure}[htbp]
  \centering
  \begin{tikzpicture}
  \begin{groupplot}[group style={group size=2 by 1, horizontal sep=1.9cm}, paperaxis, width=0.4\linewidth, xlabel={predicted $\mathbb{P}[\text{change}]$}, ylabel={observed frequency}, xmin=0, xmax=1, ymin=0, ymax=1]
    \nextgroupplot[title={GMM}, legend style={at={(1.075,-0.22)}, anchor=north, yshift=-16pt, legend columns=-1, /tikz/every even column/.append style={column sep=6pt}}]
      \addplot[black!45, dashed, forget plot] coordinates {(0,0) (1,1)};
      \addplot[color=refGray, mark=o] table[x=pred, y=obs]{\datpath reliability_gmm_raw.dat};
      \addplot[color=viridisBlue, mark=triangle*] table[x=pred, y=obs]{\datpath reliability_gmm_isotonic.dat};
      \addplot[color=viridisGreen, mark=diamond*] table[x=pred, y=obs]{\datpath reliability_gmm_beta.dat};
      \addplot[color=matchingRed, mark=square*] table[x=pred, y=obs]{\datpath reliability_gmm_va.dat};
      \legend{raw belief, isotonic, beta, Venn--ABERS}
    \nextgroupplot[title={GBM}, ylabel={}]
      \addplot[black!45, dashed, forget plot] coordinates {(0,0) (1,1)};
      \addplot[color=refGray, mark=o] table[x=pred, y=obs]{\datpath reliability_gbm_raw.dat};
      \addplot[color=viridisBlue, mark=triangle*] table[x=pred, y=obs]{\datpath reliability_gbm_isotonic.dat};
      \addplot[color=viridisGreen, mark=diamond*] table[x=pred, y=obs]{\datpath reliability_gbm_beta.dat};
      \addplot[color=matchingRed, mark=square*] table[x=pred, y=obs]{\datpath reliability_gbm_va.dat};
  \end{groupplot}
  \end{tikzpicture}
  \caption{Reliability of the filtered belief's change probability for~GMM and~GBM in terms of observed lane-change frequency versus predicted~$\mathbb{P}[\text{change}]$ for the raw belief and calibrated probabilities using isotonic regression, beta calibration, and Venn--ABERS. Bins are chosen as \emph{equal-mass} (10~quantiles), such that each marker represents the same number of test cases.}
  \label{fig:reliability}
\end{figure}

In~\Cref{fig:reliability} we show in more detail the calibration error for different bins, i.e.~the difference between average accuracy and average confidence in each bin. 
Also for this experiment, we focus on the binary problem lane-keeping versus lane-changing.
The plot makes two things evident: the benefit of calibration methods, and also that, even if the~ECE$_{\mathrm{lc}}$ is relatively small, it could happen, as shown in the~GBM plot, that for some probability ranges the gap between accuracy and confidence is large.
This is apparent in the gray curve for probabilities between~0.2 and~0.4, where the curve lies well above the diagonal.
\par
Finally, in~\Cref{tab:calibration} we also verify that calibration methods do not reduce the accuracy of the models.
This is of course a mandatory requirement: a model that reduces the calibration error but also reduces substantially the overall accuracy is not useful.
In the~GMM setting, calibration methods reduce the accuracy by between~$0.4\%$ and~$0.6\%$, while in the~GBM setting calibration even improves the accuracy in most cases, while also reducing the~ECE$_{\mathrm{lc}}$ and the ECE$_{\mathrm{cw}}$.
These results provide further empirical evidence that calibration should be regarded as an integral part of the model deployment pipeline, rather than an optional refinement, particularly in high-stakes applications such as autonomous driving, where both accuracy and reliability are essential.
\begin{table}[htbp]
  \centering
  \begin{tabular}{llcccc}
    \toprule
    Model & Calibration & $\mathrm{bAcc}$ & $\mathrm{ECE}_\mathrm{lc}$ & $\mathrm{ECE}_\mathrm{cw}$ & $\mathrm{VAw}_\mathrm{lc}$ \\
    \midrule
    GMM & raw & 89.8 & 0.093 & 0.067 & -- \\
     & isotonic & 89.4 & 0.036 & 0.029 & -- \\
     & beta & 89.2 & 0.054 & 0.046 & -- \\
     & Venn--ABERS & 89.4 & 0.041 & 0.038 & 0.033 \\
    GBM & raw & 94.4 & 0.056 & 0.037 & -- \\
     & isotonic & 95.0 & 0.022 & 0.021 & -- \\
     & beta & 94.9 & 0.031 & 0.029 & -- \\
     & Venn--ABERS & 95.2 & 0.034 & 0.035 & 0.031 \\
    \bottomrule
  \end{tabular}
  \caption{Belief calibration results (accuracy, expected calibration error, and Venn--ABERS interval width) for the raw beliefs compared to isotonic regression, beta calibration, and Venn--ABERS.}
  \label{tab:calibration}
\end{table}

\subsubsection{Runtime for training, inference and calibration}
In~\Cref{tab:runtime} we report training and inference times for the recursive Bayesian filter approach with~GMM and~GBM as emission models as well as the stacked-GBM method.
Moreover, we analyze the time required for calibration and prediction when using the Venn--ABERS predictor.
The runtime for isotonic regression and beta calibration are negligibly small and therefore not reported here.
\par
The training times for all methods are below~2\,s, which would even allow for adaptive training during deployment.
The inference times per frame (at a frame rate of~25 frames per second) are of order~$10^{-2}$ milliseconds for the prediction of~filter-GMM, filter-GBM and~stacked-GBM.
The predictions can therefore be performed in real-time and allow for instantaneous reaction based on the predicted behavior.
Also the calibration can be performed in under~2\,ms per frame.
\begin{table}[htbp]
  \centering
  \begin{tabular}{lcc}
    \toprule
    Stage & Train (s) & Inference (ms/frame) \\
    \midrule
    filter-GMM & 0.40 & 0.0055 \\
    filter-GBM & 1.63 & 0.0172 \\
    stacked-GBM & 1.43 & 0.0293 \\
    Venn--ABERS (calibration+prediction) & -- & 1.1472 \\
    \bottomrule
  \end{tabular}
  \caption{Training and inference runtime for the~filter-GMM, filter-GBM, and~stacked-GBM methods as well as calibration and prediction time for Venn--ABERS.}
  \label{tab:runtime}
\end{table}

\subsubsection{Comparison to other methods}
We first evaluate the performance of the recursive filtering-based approaches when compared to the lateral-velocity threshold, per-frame~GBM, and stacked-GBM (at window lengths of~$0.4$\,s and~$1.6$\,s), see~\Cref{sec:baseline-methods}.
For prediction horizons of at least~$h=3$\,s, the lateral-velocity threshold leads to the lowest accuracy compared to all other methods.
The accuracy gap rises with increasing prediction horizon.
The lateral-velocity threshold is not affected by surrounding observation dropout, which leads to a constant accuracy independent of the dropout rate.
The filter-GMM approach is slightly worse than the~GBM-based methods.
All~GBM methods perform similarly well for the different prediction horizons and dropout rates.
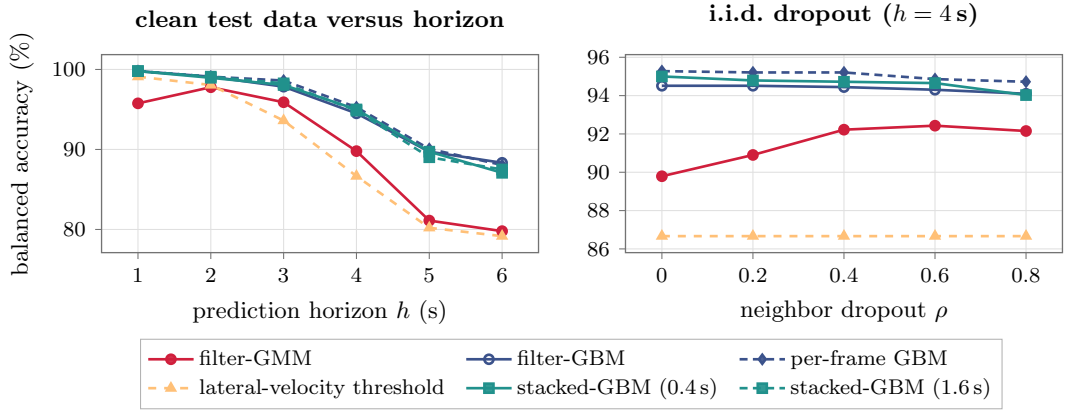
\begin{figure}[htbp]
  \centering
  \begin{tikzpicture}
  \begin{groupplot}[group style={group size=2 by 1, horizontal sep=1.15cm}, paperaxis, ylabel={balanced accuracy (\%)}, ]
    \nextgroupplot[title={clean test data versus horizon}, xlabel={prediction horizon $h$ (s)}, xtick={1,2,3,4,5,6}, legend style={at={(1.075,-0.22)}, anchor=north, yshift=-16pt, legend columns=3, /tikz/every even column/.append style={column sep=6pt}}]
      \addplot[filterGMM] table[x=H, y=filterGMM]{\datpath baselines_clean.dat};
      \addplot[filterGBM] table[x=H, y=filterGBM]{\datpath baselines_clean.dat};
      \addplot[perFrameGBM] table[x=H, y=perFrameGBM]{\datpath baselines_clean.dat};
      \addplot[rule] table[x=H, y=rule]{\datpath baselines_clean.dat};
      \addplot[stackShort] table[x=H, y=stackShort]{\datpath baselines_clean.dat};
      \addplot[stackLong] table[x=H, y=stackLong]{\datpath baselines_clean.dat};
      \legend{filter-GMM, filter-GBM, per-frame GBM, lateral-velocity threshold, stacked-GBM (0.4\,s), stacked-GBM (1.6\,s)}
    \nextgroupplot[title={i.i.d.\ dropout ($h=4$\,s)}, xlabel={neighbor dropout $\rho$}, ylabel={}]
      \addplot[rule] table[x=rho, y=rule]{\datpath baselines_dropout.dat};
      \addplot[filterGmmIid] table[x=rho, y=filterGmmIid]{\datpath baselines_dropout.dat};
      \addplot[filterGbmIid] table[x=rho, y=filterGbmIid]{\datpath baselines_dropout.dat};
      \addplot[perFrameGbmIid] table[x=rho, y=perFrameGbmIid]{\datpath baselines_dropout.dat};
      \addplot[stackIid] table[x=rho, y=stackIid]{\datpath baselines_dropout.dat};
  \end{groupplot}
  \end{tikzpicture}
  \caption{Recursive-filter methods (filter-GMM and filter-GBM) versus the external baselines (lateral-velocity threshold, per-frame GBM, and stacked-GBM using different window lengths) in terms of balanced accuracy on clean data (left) and under i.i.d.\ neighbor dropout (right).}
  \label{fig:baselines}
\end{figure}
\par
In order to compare our approach to the deep learning reference from~\cite{Cristofaro2026}, referred to as~TN2 and briefly summarized in~\Cref{sec:baseline-methods}, we now focus on clean test data.
However, as it turned out that the~GBM-based methods benefit from augmented training data even for clean test data, we train these methods on augmented data as well.
We nevertheless mention at this point that the test data used in our numerical study most likely differs from the one employed in~\cite{Cristofaro2026} due to random selection of test data.
The sampling strategy reported in~\cite{Cristofaro2026} for generating the test set was replicated in our experiments though, such that the numbers shown here are comparable and representative.
\par
\Cref{tab:benchmark} summarizes the overall accuracy (for all remaining experiments we used the balanced accuracy) of the~GMM- and~GBM-based methods over different prediction horizons~$h$ and compares them to the~TN2 method (for which only results with~$h\geq 3$ were reported in~\cite{Cristofaro2026}).
We observe that the~GMM approach is similarly accurate than~TN2 for~$h=3$ and~$h=4$.
However, for~$h=5$ and~$h=6$, the~TN2 method performs significantly better.
The~GBM emission model reaches higher accuracy for all considered prediction horizon lengths.
Together with the temporal consistency (which was not evaluated for~TN2 though), these results suggest that the~GBM approach is a valuable alternative to the~TN2 deep learning baseline that even reaches improved accuracy.
\begin{table}[htbp]
  \centering
  \begin{tabular}{cccc}
    \toprule
    $h$ (s) & GMM & GBM & TN2\,\cite{Cristofaro2026} \\
    \midrule
    1 & 95.1 & 99.5 & -- \\
    2 & 98.0 & 99.2 & -- \\
    3 & 96.6 & 98.6 & 96.7 \\
    4 & 91.7 & 95.2 & 92.5 \\
    5 & 80.0 & 91.8 & 87.8 \\
    6 & 80.6 & 89.2 & 83.9 \\
    \bottomrule
  \end{tabular}
  \caption{Overall accuracy in percent compared to the~TN2 transformer baseline from~\cite{Cristofaro2026} across different prediction horizons. For comparability with the number reported in~\cite{Cristofaro2026}, accuracy here is measured on a keep-vs-change-balanced test set which matches their construction of the test data.}
  \label{tab:benchmark}
\end{table}

\section{Discussion and outlook}\label{sec:discussion-outlook}
In this section we summarize advantages and limitations of the proposed approach, based on the numerical results from~\Cref{sec:experiments}.
Moreover, we provide an outlook to research that might follow the present paper.
\paragraph{Advantages of the method.}
First of all, we would like to highlight that the recursive Bayesian filter can in principle be applied for any length of observation window and is thus not restricted to any ``context size''.
This allows the approach to conserve and retain important information over long time horizons.
In practice, this might be beneficial if sensor breakdown occurs frequently and over longer periods of time.
The recursive Bayesian filter would be able to compensate for such missing information in a natural way.
In particular the~GMM approach provides a direct treatment of missing observations:
Instead of evaluating the~GMM for the entire observation vector or imputing data, the missing components can be marginalized out.
This is a straightforward way of handling corrupted data by simply assuming that no information is available.
For the scaled-likelihood trick and the~GBM, this marginalization is not possible.
However, as our numerical experiments showed, imputing the missing observations using approaches such as~LOCF leads to reasonable results.
\par
According to our numerical results, our method is on par with different standard baseline approaches and might show preferable behavior in terms of temporal consistency and commitment to predictions, which we assume to be important for practitioners in the context of autonomous driving.
The method provides improved temporal consistency compared to other techniques in the sense that less unexpected class flips are predicted by the model.
The temporal consistency is intrinsic to the recursive Bayesian filter since it performs a smoothing over time.
\par
We further mention that the model provides a clear probabilistic interpretation that allows for explainable intention prediction without a black-box machine learning model in the background.
Since the model provides probabilities as output, it stands to reason to also perform calibration of the model's output using for instance isotonic regression, beta calibration or~Venn--ABERS predictors as discussed in~\Cref{sec:venn-ABERS} and shown in practice in~\Cref{sec:calibration-results}.
\paragraph{Limitations of the method.}
The recursive Bayesian filtering approach nevertheless shows also limitations that we would like to discuss in the following.
First of all, as observed in our experimental study in~\Cref{sec:experiments}, the algorithm does not beat state-of-the-art methods in general.
The performance is further depending in particular on the choice of the prediction horizon and the emission model.
Marginalization of missing observations as performed for the~GMM-based emission model is competitive but not superior.
In particular for long prediction horizons, the performance of the~GMM might fall off.
A suitable technique for imputation makes the discriminative~GBM-based approaches perform similarly well, if not better in certain cases.
Also a stacked~GBM variant, see~\Cref{fig:baselines} reaches similar or even higher accuracy.
\par
Regarding calibration, we considered only three methods and a single evaluation metric (ECE). It would be beneficial to extend this analysis by incorporating additional calibration approaches, as well as alternative measures for quantifying miscalibration and unreliability in such models.
We also remark that the i.i.d.\ assumptions on the data, required by Venn–ABERS to satisfy its theoretical guarantees, are not met in our setting. An interesting direction for future work would be to derive theoretical guarantees for Venn–ABERS (or related methods) under these more realistic, non-i.i.d.\ conditions.
Similarly, the modeling assumptions in~\eqref{equ:assumptions} are relatively strong and will not be fulfilled exactly in practice.
\paragraph{Outlook and future research.}
The methods and results developed in this paper allow for further research in different directions:
In our study we focused on~GMMs and~GBMs as emission models.
Certainly, the presented recursive Bayesian filter is not restricted to those approaches and it would be of interest to also consider other methods instead.
The interpretation as hidden Markov model might further be used to derive theoretical results for the recursive filtering.
In addition, investigating additional approaches for the handling of missing observations would be a natural future research direction.
Following the experiments presented in~\Cref{sec:experiments}, we suggest to perform additional hyperparameter studies for the~GMMs and~GBMs in order to analyze the sensitivity of the approaches with respect to model parameters.
Moreover, it would be of interest to employ the described techniques in embedded devices and real-life scenarios on highways to assess their practical performance.

\section*{Statements and declarations}
\paragraph{Acknowledgements}
The authors would like to thank Martin Holler for fruitful discussions and advice.

\paragraph{Funding}
This article is based upon work from COST Action InterCoML, CA24136, supported by COST (European Cooperation in Science and Technology).

\paragraph{Code availability}
The source code used to perform the experiments shown in this paper is available in~\cite{sourcecode}.

\appendix

\section{Statistical metrics}\label{sec:statistical-metrics}
To comprehensively evaluate the performance of the proposed models, several key metrics are employed that we list and describe in the following.

\subsection{Classification metrics}
\begin{itemize}
    \item \textbf{Accuracy:}
    The accuracy describes the number of correctly classified observations and is defined as
    \begin{align*}
        \mathrm{Acc}=\frac{1}{N}\sum\limits_{i=1}^{N} \mathbf{1}_{\hat y_i=y_i},
    \end{align*}
    where~$N$ denotes the number of test instances, $y_1,\ldots,y_N\in\{\mathrm{K},\mathrm{L},\mathrm{R}\}$ are the true labels and $\hat{y}_1,\ldots,\hat{y}_N\in\{\mathrm{K},\mathrm{L},\mathrm{R}\}$ the predictions of the considered method.
    To allow for comparison with results reported in~\cite{Cristofaro2026}, we consider a balanced test set containing the same number of keep-instances as change-instances.
    \item \textbf{Recall (per class):}
    The recall or \textit{sensitivity} indicates how well the model identifies true positive events within a given intention class and is defined as
    \begin{align*}
        \mathrm{rec}_c=\frac{\mathrm{TP}_c}{\mathrm{TP}_c+\mathrm{FN}_c},
    \end{align*}
    where~$\mathrm{TP}_c$ is the number of instances correctly classified as class~$c$ and~$\mathrm{FN}_c$ is the number of instances that are actually in class~$c$ but were classified differently by the method under consideration.
    It is calculated as the proportion of true positive predictions relative to the total number of actual events of a class, which includes true positives~(TP) and false negatives~(FN).
    A high recall value signifies that the model successfully identifies most relevant events of that class.
    \item \textbf{Balanced accuracy:}
    Besides the accuracy~$\mathrm{Acc}$ introduced above, we mainly consider the balanced accuracy, which is given as the mean per-class recall and takes the number of occurrences of the individual classes into account.
    The balanced accuracy is given as
    \begin{align*}
        \mathrm{bAcc}=\frac{1}{3}(\mathrm{rec}_\mathrm{K}+\mathrm{rec}_\mathrm{L}+\mathrm{rec}_\mathrm{R}).
    \end{align*}
    Since the lane-change classes are underrepresented in the~highD dataset, we report the balanced accuracy in all figures and tables below, except when comparing to the results from~\cite{Cristofaro2026} in~\Cref{tab:benchmark}.
    \item \textbf{Precision (per class):}
    This metric measures the proportion of true positive classifications among all positive predictions made by the model and is defined as
    \begin{align*}
        \mathrm{pr}_c=\frac{\mathrm{TP}_c}{\mathrm{TP}_c+\mathrm{FP}_c}
    \end{align*}
    for class~$c\in\{\mathrm{K},\mathrm{L},\mathrm{R}\}$.
    Here, TP \textit{(true positives)} refers to the number of events correctly predicted as lane keeping~(K), left lane change~(L) or right lane change~(R), while~FP \textit{(false positives)} denotes the number of events from the other two classes incorrectly predicted as that class.
    A high precision value indicates that the model is highly reliable when predicting a class.
    In other words, the model rarely issues false alarms.
    \item \textbf{$F_1$ score (per class):}
    The~$F_1$ score for class~$c\in\{\mathrm{K},\mathrm{L},\mathrm{R}\}$ is defined as the harmonic mean of precision and recall of the respective class, i.e.
    \begin{align*}
        F_{1,c}=\frac{2\cdot\mathrm{pr}_c\cdot\mathrm{rec}_c}{\mathrm{pr}_c+\mathrm{rec}_c}.
    \end{align*}
    \item \textbf{Macro-$F_1$ score:}
    The per-class~$F_1$ scores can be aggregated into the macro-$F_1$ score by taking their mean as
    \begin{align*}
        F_1^{\mathrm{macro}}=\frac{1}{3}\sum\limits_{c\in\{\mathrm{K},\mathrm{L},\mathrm{R}\}} F_{1,c}.
    \end{align*}
\end{itemize}

\subsection{Calibration metrics}
\begin{itemize}
    \item \textbf{Expected calibration error (ECE) for lane changing:}
    For each testing instance~$i$, a predicted change probability~$p_i$ and the binary outcome~$y_i=\mathbf{1}_{\text{vehicle }i\text{ changes lane}}$ are given.
    To compute the ECE, the interval~$[0,1]$ is divided into~$B=10$ bins.
    Within bin~$b$, we collect all instances~$i$ whose predicted probability~$p_i$ falls into it, defining the set
    \begin{align*}
        \mathcal{B}_b=\left\{i\in\{1,\ldots,N\}:p_i\in\left(\frac{b-1}{B},\frac{b}{B}\right]\right\}.
    \end{align*}
    For bin~$b$, the mean predicted probability (confidence) and the observed change frequency are defined as
    \begin{align*}
        \mathrm{conf}_b=\frac{1}{|\mathcal B_b|}\sum\limits_{i\in\mathcal{B}_b} p_i
        \qquad\text{and}\qquad
        \mathrm{freq}_b=\frac{1}{|\mathcal B_b|}\sum\limits_{i\in\mathcal{B}_b} y_i.
    \end{align*}
    A perfectly calibrated model satisfies~$\mathrm{conf}_b=\mathrm{freq}_b$ for all~$b=1,\ldots,B$.
    The ECE is then defined as the average discrepancy between confidence and observed frequency, weighted by the bin size, i.e.,
    \begin{align*}
        \mathrm{ECE}_\mathrm{lc} = \sum\limits_{b=1}^{B}\frac{|\mathcal B_b|}{N}\big|\mathrm{conf}_b-\mathrm{freq}_b\big|.
    \end{align*}
    \item \textbf{Class-wise expected calibration error:}
    The class-wise ECE reports the mean one-vs-rest ECE over the three classes and is defined as
    \begin{align*}
        \mathrm{ECE}_{\mathrm{cw}} = \frac{1}{3}\sum\limits_{s\in\{\mathrm{K},\mathrm{L},\mathrm{R}\}} \mathrm{ECE}_s,
    \end{align*}
    where~$\mathrm{ECE}_s$ is the ECE associated with the predicted probability for class~$s$, computed analogously to~$\mathrm{ECE}_\mathrm{lc}$ above but using the one-vs-rest probability for class~$s$ in place of the lane-change probability~$p_i$.
    \item \textbf{Venn--ABERS interval width:}
    The Venn--ABERS predictor provides a probability interval that can be used in order to derive second-order uncertainty measures, such as the mean width of the probability interval, that is
    \begin{align*}
        \mathrm{VAw}_\mathrm{lc} = \frac{1}{N}\sum\limits_{i=1}^{N}p_{i,\mathrm{VA}}^{1}-p_{i,\mathrm{VA}}^{0},
    \end{align*}
    where~$p_{i,\mathrm{VA}}^{0}$ and~$p_{i,\mathrm{VA}}^{1}$ denote the lower and upper probability interval bounds for instance~$i$, respectively.
    Here, we focus on the mean interval width for the change probability.
\end{itemize}

\subsection{Temporal consistency and anticipation metrics}
\begin{itemize}
    \item \textbf{Flip-rate:}
    For a sequence of predicted labels~$\hat{y}_1,\ldots,\hat{y}_T$ on a track, we consider
    \begin{align*}
        \frac{1}{T-1}\sum\limits_{t=2}^{T}\mathbf{1}_{\hat{y}_t\ne\hat{y}_{t-1}},
    \end{align*}
    averaged over the testing tracks, as the flip-rate (reported below in percent).
    The flip-rate counts the number of changes of the predicted label in consecutive frames and normalizes this quantity by the total number of frames.
    A lower flip-rate corresponds to temporally steadier responses.
    \item \textbf{False-alarm rate:}
    Let~$\mathcal{T}_{\mathrm{K}}$ denote the set of lane-keeping tracks in the dataset and for a track~$\kappa\in\mathcal{T}_{\mathrm{K}}$ let~$t\in\kappa$ index its frames.
    Define for each track~$\kappa$ the highest change probability~$m_\kappa\in[0,1]$ predicted by the model as
    \begin{align*}
        m_\kappa \coloneqq \max_{t\in\kappa}\max(\mathbb{P}_t[\mathrm{L}],\mathbb{P}_t[\mathrm{R}]),
    \end{align*}
    where~$\mathbb{P}_t[c]$ denotes the predicted probability for class~$c$ at time~$t$.
    A keep track is counted as a false alarm if~$m_\kappa \geq \tau$, so the false-alarm rate at threshold~$\tau\geq 0$ is given by
    \begin{align*}
        \mathrm{FAR}(\tau) = \frac{\bigl|\{\kappa\in\mathcal{T}_{\mathrm{K}} : m_\kappa \geq \tau\}\bigr|}{|\mathcal{T}_{\mathrm{K}}|},
    \end{align*}
    i.e.~the number of lane-keeping tracks classified as lane change by the model divided by the total number of lane-keeping tracks.
    We fix the threshold~$\tau$ by requiring a target rate of~$5\%$ false alarms, which means that~$\tau$ is chosen as the smallest value such that~$\mathrm{FAR}(\tau)\leq 5\%$.
    The threshold~$\tau$ can then be used to classify a track as not being lane-keeping if there is at least one frame~$t$ such that either the probability of a lane change to the left or to the right is at least~$\tau$.
    The false-alarm rate is thus determined in such a way that only~$5\%$ of the lane keeping tracks are labeled incorrectly.
    \item \textbf{Detection lead time:}
    Let~$\tau$ denote the false-alarm threshold from the definition of the false-alarm rate above; it is a single, track-level threshold on~$\max(\mathbb{P}_t[\mathrm{L}],\mathbb{P}_t[\mathrm{R}])$ that is shared across both sides.
    For a lane change to side~$c\in\{\mathrm{L},\mathrm{R}\}$, we say the model \emph{fires} at frame~$t$ if the probability of the correct side is at least the threshold, i.e.~$\mathbb{P}_t[c]\geq\tau$.
    If the actual crossing happens in frame~$s$, we consider all frames~$t<s$ before the crossing and define two notions of detection lead time:
    \begin{itemize}
        \item \emph{First-touch lead}: Consider the earliest firing frame~$t_\mathrm{ftl}$, i.e.~the smallest~$t<s$ such that~$\mathbb{P}_t[c]\geq\tau$.
        The first-touch lead time is then defined as~$s-t_\mathrm{ftl}$.
        \item \emph{Sustained lead}: Consider the earliest frame~$t_\mathrm{sl}<s$ from which~$\mathbb{P}_t[c]\geq\tau$ holds for all frames~$t$ with~$t_\mathrm{sl}\leq t<s$, i.e.~the model fires continuously up to the crossing.
        The sustained lead time is then defined as~$s-t_\mathrm{sl}$.
    \end{itemize}
    The lead times are only defined for \emph{detected} maneuvers -- those that fire at least once before the crossing (first-touch), and that additionally keep firing up to the last frame before the crossing (sustained).
    The reported numbers should therefore be interpreted alongside the recall.
    \item \textbf{Recall at required lead~$d$:} We also report the recall when a first-touch or sustained lead of at least~$d$ seconds is required, i.e.~the fraction of true lane changes whose first-touch/sustained lead is at least~$d$.
    Intuitively, this measures how many lane changes the model would catch early enough to be useful: a larger~$d$ demands earlier detection, so recall decreases as~$d$ grows.
    The rate at which the recall drops off with increasing~$d$ therefore characterizes how far in advance the model can reliably anticipate maneuvers.
    We consider different values for~$d$ in order to analyze the anticipation capabilities of the recursive Bayesian filter.
    A value of~$d=0$ recovers the plain detection recall at the fixed false-alarm rate.
\end{itemize}

\section{Hyperparameters}\label{sec:hyperparameters}
The~highD dataset was split on the track level (not per frame) into a training and a testing subset.
In total, $70\%$ of the available data ($12,546$ tracks) were used as training data.
The test set contained the remaining~$30\%$ ($5,377$ tracks) of the dataset.
Training of the different approaches is based on the training tracks only.
The test set itself consists of a calibration set and an evaluation set (both chosen as~$15\%$ of the entire data).
The evaluation set is the same across all experiments.
To allow for a fair comparison of the performance of all methods mentioned in the paper, we average the results over three independent samplings (using fixed seeds) of the evaluation data, drawing one instance per track at a random frame within~$4$s prior to a maneuver.
The number of instants per class in the evaluation set is chosen such that each class (lane keeping, left change, right change) is present in exactly the same number of tracks (roughly~$160$ lane keeping, left change, and right change instances resulting in about~$480$ evaluation samples in total).
\par
For the~GMM as emission model, we use the~\texttt{GaussianMixture} class from~\texttt{scikit-learn} with three Gaussian components per class (experiments not discussed in detail in this paper showed that adding more components does not have a significant influence on the accuracy of the resulting model), a full learnable covariance and regularization of~$10^{-2}$ added to the diagonal of the covariance.
We use the~\texttt{HistGradientBoostingClassifier} for the~GBM emission term (due to the large number of samples in the training dataset) with a maximum of~$400$ iterations of the boosting process, a learning rate of~$0.05$, the maximum number of leaf nodes in the trees set to~$31$, a minimum of~$50$ samples per leaf, and early stopping enabled based on a validation set using~$15\%$ of the samples.
We refer to the documentation of the~\texttt{scikit-learn} package for additional details.

\bibliographystyle{abbrv}
\bibliography{literature}

\end{document}